\documentclass[9pt, twocolumn]{extarticle} 
\usepackage{amsmath,amssymb,amsfonts}
\usepackage{algorithmic}
\usepackage{graphicx,color}
\usepackage{textcomp}
\usepackage{xcolor}
\usepackage{hyperref}
\hypersetup{hidelinks}
\usepackage{algorithm,algorithmic}
\def\BibTeX{{\rm B\kern-.05em{\sc i\kern-.025em b}\kern-.08em
    T\kern-.1667em\lower.7ex\hbox{E}\kern-.125emX}}
\AtBeginDocument{\definecolor{tmlcncolor}{cmyk}{0.93,0.59,0.15,0.02}\definecolor{NavyBlue}{RGB}{0,86,125}}

\usepackage[margin=0.7in]{geometry}

\usepackage[labelformat=simple]{subcaption}

\usepackage{tikz}
\usepackage{multirow}
\usepackage{makecell}
\usepackage{authblk}

\DeclareMathOperator{\dB}{dB}
\DeclareMathOperator{\PL}{PL}
\DeclareMathOperator{\power}{P}
\DeclareMathOperator{\RM}{RM}
\DeclareMathOperator{\FSPL}{FSPL}

\usepackage[giveninits=true]{biblatex}
\bibliography{bib_rmdir}

\makeatletter
\def\@maketitle{%
  \newpage
  \null
  \vskip 2em%
  \begin{center}%
  \let \footnote \thanks
    {\Huge\bfseries\@title \par}%
    \vskip 1.5em%
    {\large
      \lineskip .5em%
      \begin{tabular}[t]{c}%
        \@author
      \end{tabular}\par}%
    \vskip 1em%
    {\large \@date}%
  \end{center}%
  \par
  \vskip 1.5em}
\makeatother

\title{Radio Map Estimation - An Open Dataset with Directive Transmitter Antennas and Initial Experiments}

\author[$\star$]{Fabian Jaensch} 
\author[$\star$]{Giuseppe Caire}
\author[$\dagger$]{Begüm Demir}

\affil[$\star$]{Communications and Information Theory Group, TU Berlin}
\affil[$\dagger$]{Remote Sensing Image Analysis Group, TU Berlin}

\date{}

\begin{document}

\maketitle
\begin{abstract}
    Over the last years, several works have explored the application of deep learning algorithms to determine the large-scale signal fading (also referred to as ``path loss'') between transmitter and receiver pairs in urban communication networks.
    The central idea is to replace costly measurement campaigns, inaccurate statistical models or computationally expensive ray-tracing simulations by machine learning models which, once trained, produce accurate predictions almost instantly.
    Although the topic has attracted attention from many researchers, there are few open benchmark datasets and codebases  that would allow everyone to test and compare the developed methods and algorithms.
    We take a step towards filling this gap by releasing a publicly available dataset of simulated path loss radio maps together with realistic city maps from real-world locations and aerial images from open datasources.
    Initial experiments regarding model architectures, input feature design and estimation of radio maps from aerial images are presented and the code is made available.\footnote{\url{https://github.com/fabja19/RML}}


    \vspace{0.2cm}\small \textit{This work has been submitted to the IEEE for possible publication. Copyright may be transferred without notice, after which this version may no longer be accessible.}
\end{abstract}


\section{INTRODUCTION}\label{sec:introduction}
Wireless communication systems are fundamentally based on radio waves radiated by a transmitter (Tx) antenna carrying a signal in order to send information to a receiving (Rx) antenna.
The quality of the wireless channel in terms of the received signal strength (RSS) depends on the output power of the Tx and the signal attenuation or path loss.
This attenuation occurs due to large-scale fading effects along different paths such as distance dependent power dissipation in free space, reflections from building and street surfaces and attenuation by tree canopies. 
To capture this quantity, we consider the radio map of a transmitter consisting of the path loss values for different potential Rx locations of interest.

Important applications of radio maps include the optimization of cellular network structures to achieve a good coverage of an area \cite{coverage}, link scheduling \cite{link_scheduling}, localization of user equipment (UE) based on RSS measurements at the Rx \cite{rss_loc} or, in the context of 5G and 6G, beam management \cite{beams}.
Since measurement campaigns are impractical and too expensive on a large scale, different methods have been developed to approximate the path loss.
Statistical path loss models which express the fading solely as a function of the distance between Tx and Rx and stochastic effects are inherently not able to capture the radio wave propagation in a specific environment.
Ray-tracing simulations provide an option to approximate the underlying physical phenomena with high accuracy \cite{ray-tracing-accuracy}. 
However, due to the relatively long run times of the simulations and the need for detailed 3D city models, they can be infeasible for applications in real-time or on large-scale.

Recently, several works have explored the application of machine learning (ML) algorithms and, in particular, deep neural networks (DNNs) to the path loss estimation problem.
The general idea is that a properly designed ML model can learn the underlying physical phenomena of radio wave propagation to a certain extent, given a sufficient amount of training data.
Albeit the training process can be very time and resource intensive, the run time of the inference task once the model is deployed is typically in the order of milliseconds.
This opens up a range of applications in which statistical path loss models are too inaccurate and ray-tracing is too slow.

\subsection{PROBLEM FORMULATION}\label{sec:pathloss}
Consider a Tx-Rx pair in a fixed environment with $n$ (multi\nobreakdash-)paths establishing the link from the Tx to the Rx.
These may include a direct line-of-sight path and multipaths undergoing reflections from surfaces such as the ground or building walls and diffractions around edges or corners of objects.
In our simulation scenario described in Section \ref{sec:dataset}, we assume that the wavelength of the communication signal is significantly smaller than the accuracy of spatial positions and shapes of objects in the environment.
This implies that it is impossible to model the phase differences between different paths appropriately, and any small-scale fading due to constructive and destructive interference between paths arriving with different phase shifts is modeled as a random effect.
In fact, in standard wireless communication theory it is customary to separate the large-scale pathloss from the small-scale random fading and consider normalized statistics for the latter (e.g., Rayleigh/Rician fading with unit second moment , see \cite{molisch}). 
Mathematically, this means modeling the channel coefficient between a Tx and a Rx at fixed positions as a stationary Gaussian random process $H(t,f)$ over time $t$ and frequency $f$ with second moment $G = E[\left|H(t,f)\right|^2]$  that depends on the position of the Tx and Rx in the scattering environment.~\footnote{If directional antenna patterns are used, then $G$ depends also on the orientation of Tx and Rx.}
The so-called ``pathloss'' is defined as the inverse of the second moment of the channel, i.e., $\PL = 1/G$ and, for a fixed Tx, this is a function of the location of the Rx. 
In addition, due to the non-coherent combining of the multipath components superimposing at the Rx location, $G$ is equal to the sum of the second moments of the individual multipath components. 
From a physical viewpoint, $G$ is given by the time-frequency average of the received power over the signal bandwidth and a sufficiently long time interval to span several coherence intervals. 
While the instantaneous value of the channel coefficients $H(t,f)$ varies randomly across time and frequency,~\footnote{Although it is roughly constant across time-frequency ``tiles'' of size $T_c \times W_c$, where $T_c$ is the coherence time and $W_c$ is the coherence bandwidth, see \cite{molisch}.} its second moment $G$ is constant over time and frequency and depends only on the location. 
Consequently, in the following all power values are assumed to be averaged in time over a sufficiently large interval and integrated over the signal bandwidth, in order to obtain a meaningful representation.
We use subscript $W$ and $\dB$ to distinguish between values in Watts and decibel, respectively.

Due to the non-coherent path combining explained above, denoting as $P_i$ the power of the $i$-th path arriving at the Rx,  the \textit{received power} (in watts) is given by the sum of the individual paths powers as 
\begin{equation}
    (\power_R)_{W}   =  \sum_{i=1}^n (\power_i)_W .
\end{equation}
By conversion to dB scale and subtraction of the power radiated by the Tx, $P_T$, we obtain the \textit{path loss} (in dB)
\begin{equation}\label{eq:pathloss}
    (\PL)_{\dB}   =   (\power_R)_{\dB} - (\power_T)_{\dB}.
\end{equation}

The \textit{radio map} of the Tx for a fixed set $D\subset\mathbb{R}^2$ of locations of interest may now formally be defined as a function $\RM:D\rightarrow\mathbb{R}$, which maps each location to the path loss value for an Rx in the considered position according to \eqref{eq:pathloss}.
In the following, we will only consider the case that $D$ corresponds to a uniform grid over a square-shaped area.
This allows us to regard the radio map as a matrix or image.

As elaborated in \cite{radiounet}, it is reasonable to truncate the path loss from below at a threshold corresponding to the noise floor, as lower power levels are irrelevant in practice, see Section \ref{sec:rm_simulations}. 

\subsection{GENERAL APPROACH AND RELATED WORKS}\label{sec:related_works}
A common way to formulate the path loss is to use the log normal shadowing models in different variations (e.g. the 3GPP model described in \cite{3gpp_log_normal}).
They originate from the idea to express the path loss merely as a function of the distance between Tx and Rx and to model deviations as the realization of a Gaussian random variable in the log (dB) domain. .
Albeit the resulting graph of the function together with the stochastic spread may fit measurement points, these models completely disregard the radio wave propagation in a specific environment along specific paths.

Ray-tracing \cite{ray-tracing} provides a more accurate, site-specific solution by modeling electromagnetic waves as rays launched in a very fine subdivision of the space around the Tx.
Reflections from surfaces and potentially other effects such as diffractions or transmissions are calculated upon contact with objects \cite{ray-tracing-accuracy}.
The rays arriving at the Rx are combined to generate the path loss defined in \eqref{eq:pathloss} or other channel state information.
As mentioned before, potential drawbacks for certain applications are the runtime of typically at least several seconds to generate a complete radio map and the need to have a complete 3D model of the environment available.

Other methods include completing the radio map from sparse measurements \cite{rm_completion} or estimating the path loss for a single Rx position at a time via ML \cite{ml_single}.

In the following, we describe approaches to predict the whole radio map at once using ML.
Each pixel in the target radio map, represented by a two-dimensional image, corresponds to a different location in the considered environment from a birds-eye perspective, and its value is the path loss in an appropriate scale.
The input information about the presence and potentially height of objects like buildings or trees, the transmitter position and possibly other parameters that differ from sample to sample, are usually encoded in two-dimensional images of the same shape as the target and stacked along an additional channel dimension.
All works described in the following build on convolutional neural network (CNN) models \cite{cnn}, which have been shown to be effective in other tasks such as semantic segmentation, image denoising or image-to-image translation that are structurally similar in the sense that input and output of the model are image-like tensors. 

A first seminal contribution investigating the prediction of the whole radio map at once using CNNs came out of our research group \cite{radiounet} and considered a UNet \cite{unet} model trained in a supervised manner to approximate radio maps generated with two-dimensional ray-tracing from city maps and Tx locations.
Presence or absence of buildings and the Tx are encoded in binary input images.
The authors explore strategies to incorporate available signal measurements and adapt the trained models to more realistic scenarios and show applications to coverage estimation and fingerprint localization.
Several other works have extended this approach to the more challenging scenario of radio maps generated with 3D ray-tracing, e.g. \cite{plnet}, \cite{fadenet}, \cite{radiotrans}.
In these works, the binary input information per pixel is usually replaced by height information.
Some works have  experimented with more variables between the different samples encoded in additional inputs, for example the carrier frequency of the transmitted signal \cite{radiotrans} or antenna patterns and orientations \cite{plnet}.
Albeit the basic encoder-decoder with skip-connections design of the UNet has been used in most works since then, it has been soon recognized that standard UNets are inherently limited when it comes to propagating information over longer distances, which is especially important to accurately predict reflections.
Several approaches make use of dilated convolutions \cite{dilated} to increase the receptive field \cite{ziemann21}, \cite{deepray}, \cite{pmnet}, whereas in \cite{radiotrans} vision transformer layers \cite{vit} are added to the CNN to allow modeling long-range relationships.
Besides changes to the network architecture, some authors propose feature engineering to resolve the described problem and improve the accuracy.
In \cite{qiu22}, the network is provided with an input image containing the spatial distance of each pixel position to the transmitter location.
The authors of \cite{radiotrans} propose to feed the coordinates of each pixel and the Tx to the network in all positions.
To show the validity of training CNNs on simulated data, in \cite{plnet} a model tested first on radio maps generated by ray-tracing is retrained on real world measurements and shown to perform better than all conventional methods.
Some papers have also considered estimation of radio maps of Tx mounted on unmanned aerial vehicle (UAV) from building height maps \cite{vaegan} or satellite images \cite{plgan}.
In this scenario, the complex propagation phenomena between buildings do not arise to the same extent.
These radio maps therefore pose a very different, arguably a lot easier, problem.
    
Unfortunately, in most cases code and datasets are not published, which makes it difficult or impossible to reproduce and inspect results.
A comparison of the proposed approaches solely based on numbers presented in the papers is meaningless, since every dataset with different characteristics, e.g. the underlying propagation model and software, 2D or 3D, Tx and Rx properties etc., potentially poses a completely different problem.
In some cases the reason could be that city models needed to simulate the ground truth radio maps are difficult to find or to generate in an appropriate format, therefore a few works have relied on proprietary commercial data \cite{fadenet}, \cite{plgan}.
The only open datasets we are aware of are the one presented in \cite{radiounet}, an extension thereof to 3D with random building heights \cite{cagkan3d}, and another 2D dataset from \cite{pmnet}.
Albeit these datasets and the corresponding articles have contributed with valuable investigations to the are of radio map prediction, we have identified a few limitations and possible extensions:
None of them is based on 3D urban environments from the real world, including buildings with realistic heights and featuring trees and other vegetation.
Two datasets only consider 2D environments and propagation, which poses a significantly less complex problem than 3D propagation.
Antennas are assumed to be perfectly isotropic in all of them, which is not possible in practice and leaves open the question how to adapt the proposed methods to higher frequency bands, that require focusing the power radiated by the Tx.
Furthermore, for applications in areas in which the environment geometry is not or not exactly know, it would be desirable to predict the radio map from a wider available data source such as images.

\begin{table*}
    \small
    \begin{tabular}{|c||c|c|c|c|}
        \hline
        Dataset         &   RadioMapSeer (RadioUNet \cite{radiounet}) &   RadioMapSeer3D \cite{cagkan3d} & USC (PMNet \cite{pmnet}) & RMDirectionalBerlin (ours) \\
        \hline
        \hline
        Dimension       &   2D      &       3D      &       2D      &   3D  \\
        \hline
        Environment data&   buildings &   \makecell[tc]{buildings with a single\\ random height value}  &   buildings &   building and vegetation nDSMs   \\            
        \hline
        Antennas (Tx)   &   isotropic   &   isotropic   &   isotropic   &   directional \\
        \hline
        \#Samples       &   $56080$ &  $50680$  &   $19016$ &   $74515$     \\
        \hline
        Additional data &   \makecell[tc]{repeated simulations with\\different propagation models\\and additional cars,\\time-of-arrival maps} & -  &   -   &   \makecell[tc]{aerial images}\\
        \hline
    \end{tabular}
    \caption{Overview of open radio map datasets.}
    \label{table:datasets}
\end{table*}

\subsection{OUR CONTRIBUTION}\label{sec:contribution}
To address the limitations of the existing approaches mentioned in Section \ref{sec:related_works}, we present a new radio map dataset.
We have generated city geometries from real places in the city of Berlin and conducted ray-tracing simulations to obtain a large collection of path loss maps modeling typical urban cellular networks, described in Section \ref{sec:dataset}.
The dataset and the code for our experiments are available via our Github page\footnote{\url{https://github.com/fabja19/RML}}.
In contrast to the other publicly available radio map datasets we are aware of \cite{radiounet}, \cite{pmnet}, \cite{cagkan3d}, ours is the first one to feature directional antennas at the Tx and to include trees and realistic building heights with approximated roof shapes taken from the real world.
We also add aerial imagery of the same locations, opening up the possibility to train models for situations where exact knowledge of the city geometry is not available.
Table \ref{table:datasets} provides an overview of the similarities and differences between other openly accessible datasets and ours.

We perform experiments with different CNN architectures and compare various ways of encoding the city geometry, the relative position of the Tx to locations on the map and the Tx antenna characteristics, including several ideas proposed in the literature (Section \ref{sec:input_features}).
Furthermore, we provide some initial study on predicting the radio map from limited information in the form of aerial images and potentially unclassified height maps.
This idea has been considered in related works on predicting the path loss for a single Rx (see e.g. \cite{img_singlerx1}, \cite{img_singlerx2}) and predicting the radio map of a Tx mounted on a UAV \cite{plgan}.
Our work is, however, the first one that investigates the significantly more challenging problem of predicting the whole radio map from images, where the RSS is at the ground (1.5m height from the ground) and the transmitting base station (BS) is mounted on a building roof.
Lastly, we propose the usage of deformable convolutional layers \cite{dcn} for the radio map prediction task and show their effectiveness.

\section{SIMULATION SETUP AND DATASET}\label{sec:dataset}
Our dataset consists of in total 74,515 radio maps simulated with ray-tracing on 424 city maps from Berlin.
The environment data and the radio maps cover an area of $256$m$\times256$m with a spatial resolution of $1$m.
For each city map, several potential Tx locations on the corners and edges of building roofs have been identified and for all of these locations, we have conducted simulations with different antenna characteristics and orientations.
\begin{figure*}[h]
    \centering
    \begin{subfigure}[t]{0.2\textwidth}
        \includegraphics[width=\textwidth]{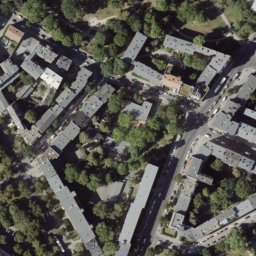}
        \caption{}
        \label{fig:maps:img}
    \end{subfigure}
    \begin{subfigure}[t]{0.2\textwidth}
        \includegraphics[width=\textwidth]{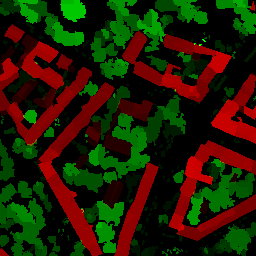}
        \caption{}
        \label{fig:maps:gis}
    \end{subfigure}
    \begin{subfigure}[t]{0.25\textwidth}
        \includegraphics[width=\textwidth]{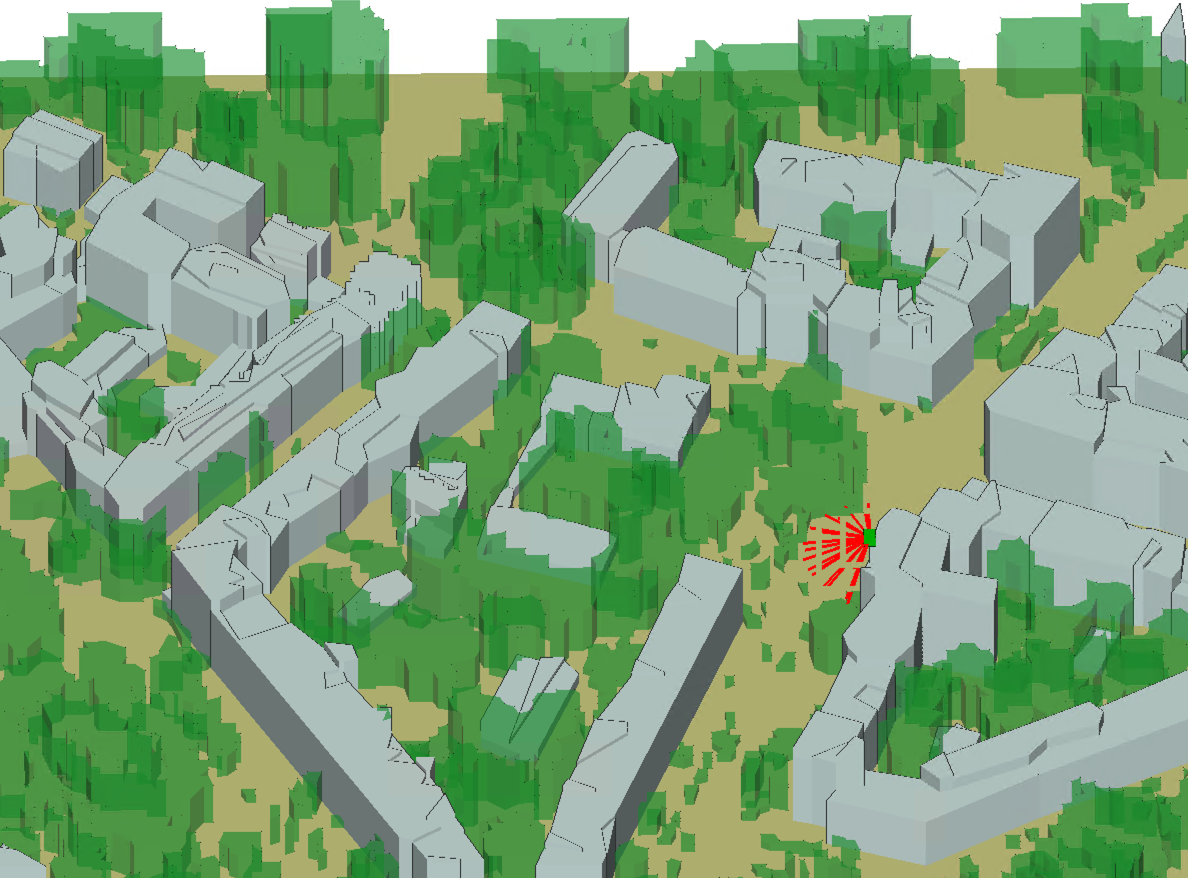}
        \caption{}
        \label{fig:maps:3d}
    \end{subfigure}
    \begin{subfigure}[t]{0.2\textwidth}
        \includegraphics[width=\textwidth]{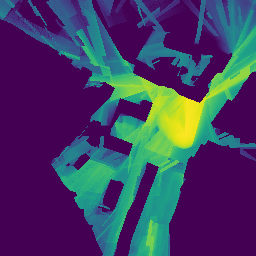}
        \caption{}
        \label{fig:maps:rm}
    \end{subfigure}
    \begin{subfigure}[t]{0.054\textwidth}
        \includegraphics[width=\textwidth]{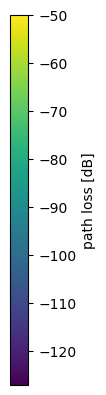}
    \end{subfigure}
    \caption{Example of sample from the dataset -- \subref{fig:maps:img} aerial image \subref{fig:maps:gis} nDSMs of buildings (red) and vegetation (green) overlayed with darker colors corresponding to larger height values \subref{fig:maps:3d}
    3D model in simulation with Tx (green cube) and different orientations of the antenna main lobe (red lines) \subref{fig:maps:rm} example of a radio map}
    \label{fig:maps}
\end{figure*}

\subsection{CITY MAPS}\label{sec:city_maps}
Our goal was to simulate cellular networks in urban environments corresponding to places in the real world and featuring buildings  with realistic shapes and heights and also trees.
Although a few public institutions provide open 3D models of certain cities, municipalities or even whole countries, we could not find any that contained information about foliage.
Available height maps or digital surface models (DSMs), on the other hand, lack the necessary classification of objects as buildings, vegetation and ground.
We have therefore decided to generate a new dataset of city maps from raw airborne \textit{light detection and ranging} (LiDAR) point clouds provided by the Geoportal Berlin \cite{geoportal_berlin}.
Using the software LAStools \cite{lastools}, we have automatically separated the point clouds into the classes ground, building and vegetation.
The recognition of buildings has been further improved by incorporating building footprints from \cite{geoportal_berlin}, categorizing all elevated (above-ground) points inside the footprints as buildings.
From there we extracted normalized digital surface models (nDSM), representing the height above the ground for the two classes building and vegetation. 
Since Berlin shows small deviation in elevation in most areas, we figured that approximating the ground as a flat surface and normalizing the heights of objects relative to it results in an acceptable simplification.
Although the resulting maps contain a few minor errors in the classification or shapes in some places, we find the resulting height maps and 3D models to provide a good approximation of reality (see Fig. \ref{fig:maps}).

\subsection{RADIO MAP SIMULATIONS}\label{sec:rm_simulations}
The ground truth radio maps were generated with the GPU accelerated X3D propagation model in the widely used ray-tracing software Wireless InSite \cite{wi}.
Since it was impossible to determine the exact materials of specific houses or types of trees, we have chosen standard material types for these upon import to the simulation software.
The nDSMs have been converted to polygons for the simulations, by grouping up neighboring pixel with approximately the same height and interpolating the boundaries.
The ground and all buildings are assumed to represent solid structures, allowing reflections from surfaces and diffractions around edges but no transmissions.
Vegetation on the other hand is modeled as a solely attenuating material.
Exploiting that the exact shape of foliage objects is therefore less important, we have chosen less accurate interpolation for the vegetation layers upon constructing the polygons.
By this, we could significantly reduce the runtime, which depends heavily on the number of faces in the environment geometry, and hence generate a larger dataset.

To simulate a realistic cellular environment, Tx that model cellular base stations were placed on the edges of buildings at a height of $2$m from the roof and restricted to heights between $6$m and $30$m above the ground. 
A dense grid of receivers with isotropic antennas and with a spacing of $1$m at a height of $1.5$m was defined to model typical user equipments (UE) in the network, e.g. smartphones of people walking on the sidewalks or devices in cars.

\begin{figure}[ht]
    \centering
    \begin{subfigure}[t]{0.14\textwidth}
    \includegraphics[width=\textwidth]{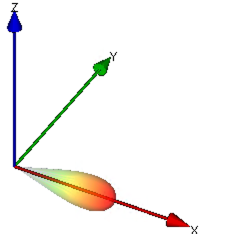}
    \caption{$(15^{\circ}, 30^{\circ})$}
    \label{fig:ant:narrow}
    \end{subfigure}
    \hfill
    \begin{subfigure}[t]{0.14\textwidth}
    \includegraphics[width=\textwidth]{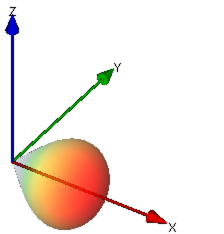}
    \caption{$(45^{\circ}, 90^{\circ})$}
    \label{fig:ant:middle}
    \end{subfigure}
    \hfill
    \begin{subfigure}[t]{0.14\textwidth}
    \includegraphics[width=\textwidth]{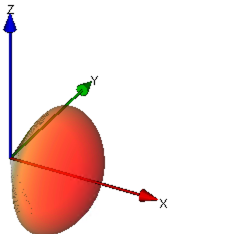}
    \caption{$(90^{\circ}, 120^{\circ})$}
    \label{fig:ant:wide}
    \end{subfigure}
    \caption{3D plot of antenna radiation patterns. The values in bracket indicate the half power beam width and first null beam width, respectively.}
    \label{fig:ant}
\end{figure}

For the Tx, we have selected the directional antenna type in the ray-tracing software.
It allows to simulate an idealized main beam without any side lobes.
By adjusting the parameters half power beam width and first null beam width, it can model very narrow beams or antennas covering wider sectors.
For each Tx position, we test several combinations of azimuth and tilt angles in the direction pointing away from the building, on which the Tx is placed on, with a line of sight algorithm.
Orientations which do not cover at least a certain area on the ground are disregarded.
For all angle combinations found, we select one of the narrow and one of the wide antenna patterns in Table \ref{table:antennas}, for separate simulations.

\begin{table*}[!ht]
    \centering
    \begin{tabular}{|c|cccc|cccc|}
        \hline
        {Pattern}                        &   \multicolumn{4}{c|}{Narrow} &   \multicolumn{4}{c|}{Wide}   \\
        \hline
        Half power beam width   &   $15^{\circ}$  &   $15^{\circ}$  &   $30^{\circ}$ &   $45^{\circ}$  &   $15^{\circ}$  &   $30^{\circ}$  &   $45^{\circ}$  &   $90^{\circ}$  \\  
        \hline
        First null beam width   &   $30^{\circ}$  &   $60^{\circ}$  &   $60^{\circ}$  &   $60^{\circ}$  &   $90^{\circ}$  &   $90^{\circ}$  &   $90^{\circ}$  &   $120^{\circ}$ \\
        \hline
    \end{tabular}
    \caption{Antenna parameters.}
    \label{table:antennas}
\end{table*}

In Table \ref{table:simulation_parameters} we list further parameters used.
\begin{table}[!ht]
    \centering
    \begin{tabular}{|c|c|}
        \hline
        Carrier frequency   &   $3.7$ GHz \\
        Maximum path loss & $-50$ dB \\
        Noise floor  &   $-127$ dB \\
        Number of reflections   &   2 \\
        Number of diffractions   &   1 \\
        Number of transmissions   &   0 \\
        \hline
    \end{tabular}
    \caption{Simulation and dataset parameters.}
    \label{table:simulation_parameters}
\end{table}
The simulation output is the received power at each Rx on the grid. 
Subtracting the input power, we obtain the path loss in dB (Section \ref{sec:pathloss}), which is then cut off at a threshold corresponding to the noise floor of $-127$dB and linearly scaled to the interval $[0,1]$ to obtain the radio map as a gray scale image.
The rescaling assures that, in contrast do $\dB$-scale, the most relevant parts of the radio map containing strong signal dominate parts with very low signal in terms of magnitude.
See \cite{radiounet} for a comprehensive explanation and motivation of the thresholding and scaling.

\subsection{OTHER DATA}\label{sec:other_data}
We also include images from \cite{geoportal_bb} taken about $2$ months after the LiDAR measurements, cut and downscaled to match the position and resolution of the nDSMs.
Furthermore, the dataset contains 2D polygons with height attributes extracted from the nDSMs, which have been used for the ray-tracing simulations.
Files containing line-of-sight information for each radio map (Section \ref{sec:los}) are also included.

\section{EXPERIMENT DESIGN}\label{sec:experiments}

\subsection{CNN-ARCHITECTURES}\label{sec:architectures}
As a lightweight baseline model we use the RadioUNet \cite{radiounet}, more concretely the first part of the WNet described by the authors.
In principle, it follows the structure of the original UNet \cite{radiounet}, but it features more down and upsampling layers and in some parts convolutions with a larger kernel size allowing to propagate information over longer distances.
Furthermore, we include experiments with the PMNet proposed in \cite{pmnet}, featuring a relatively deep encoder consisting of stacked ResNet-layers and several parallel convolutional layers with varying \textit{dilation} \cite{dilated} after the encoder.
These were the only architectures designed for the radio map prediction task we could find that have been made publicly available.

\tikzset{every picture/.style={line width=0.75pt}} 
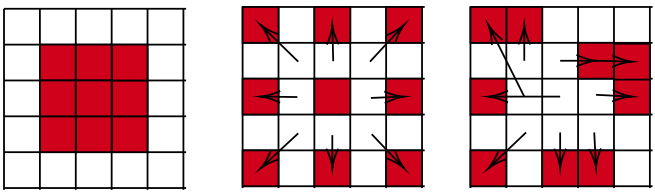
\begin{figure}[h]
    \centering
    \resizebox{\linewidth}{!}{
    \begin{tikzpicture}[x=0.75pt,y=0.75pt,yscale=-1,xscale=1]

    \draw  [draw opacity=0] (111,98) -- (212.5,98) -- (212.5,199) -- (111,199) -- cycle ; \draw   (111,98) -- (111,199)(131,98) -- (131,199)(151,98) -- (151,199)(171,98) -- (171,199)(191,98) -- (191,199)(211,98) -- (211,199) ; \draw   (111,98) -- (212.5,98)(111,118) -- (212.5,118)(111,138) -- (212.5,138)(111,158) -- (212.5,158)(111,178) -- (212.5,178)(111,198) -- (212.5,198) ; \draw    ;
    \draw  [draw opacity=0] (244,97) -- (345.5,97) -- (345.5,198) -- (244,198) -- cycle ; \draw   (244,97) -- (244,198)(264,97) -- (264,198)(284,97) -- (284,198)(304,97) -- (304,198)(324,97) -- (324,198)(344,97) -- (344,198) ; \draw   (244,97) -- (345.5,97)(244,117) -- (345.5,117)(244,137) -- (345.5,137)(244,157) -- (345.5,157)(244,177) -- (345.5,177)(244,197) -- (345.5,197) ; \draw    ;
    \draw  [draw opacity=0] (371,97) -- (472.5,97) -- (472.5,198) -- (371,198) -- cycle ; \draw   (371,97) -- (371,198)(391,97) -- (391,198)(411,97) -- (411,198)(431,97) -- (431,198)(451,97) -- (451,198)(471,97) -- (471,198) ; \draw   (371,97) -- (472.5,97)(371,117) -- (472.5,117)(371,137) -- (472.5,137)(371,157) -- (472.5,157)(371,177) -- (472.5,177)(371,197) -- (472.5,197) ; \draw    ;
    \draw  [fill={rgb, 255:red, 208; green, 2; blue, 27 }  ,fill opacity=1 ] (131,118) -- (151,118) -- (151,138) -- (131,138) -- cycle ;
    \draw  [fill={rgb, 255:red, 208; green, 2; blue, 27 }  ,fill opacity=1 ] (244,137) -- (264,137) -- (264,157) -- (244,157) -- cycle ;
    \draw  [fill={rgb, 255:red, 208; green, 2; blue, 27 }  ,fill opacity=1 ] (284,97) -- (304,97) -- (304,117) -- (284,117) -- cycle ;
    \draw  [fill={rgb, 255:red, 208; green, 2; blue, 27 }  ,fill opacity=1 ] (324,137) -- (344,137) -- (344,157) -- (324,157) -- cycle ;
    \draw  [fill={rgb, 255:red, 208; green, 2; blue, 27 }  ,fill opacity=1 ] (284,137) -- (304,137) -- (304,157) -- (284,157) -- cycle ;
    \draw  [fill={rgb, 255:red, 208; green, 2; blue, 27 }  ,fill opacity=1 ] (171,158) -- (191,158) -- (191,178) -- (171,178) -- cycle ;
    \draw  [fill={rgb, 255:red, 208; green, 2; blue, 27 }  ,fill opacity=1 ] (171,138) -- (191,138) -- (191,158) -- (171,158) -- cycle ;
    \draw  [fill={rgb, 255:red, 208; green, 2; blue, 27 }  ,fill opacity=1 ] (151,158) -- (171,158) -- (171,178) -- (151,178) -- cycle ;
    \draw  [fill={rgb, 255:red, 208; green, 2; blue, 27 }  ,fill opacity=1 ] (131,158) -- (151,158) -- (151,178) -- (131,178) -- cycle ;
    \draw  [fill={rgb, 255:red, 208; green, 2; blue, 27 }  ,fill opacity=1 ] (131,138) -- (151,138) -- (151,158) -- (131,158) -- cycle ;
    \draw  [fill={rgb, 255:red, 208; green, 2; blue, 27 }  ,fill opacity=1 ] (171,118) -- (191,118) -- (191,138) -- (171,138) -- cycle ;
    \draw  [fill={rgb, 255:red, 208; green, 2; blue, 27 }  ,fill opacity=1 ] (151,118) -- (171,118) -- (171,138) -- (151,138) -- cycle ;
    \draw  [fill={rgb, 255:red, 208; green, 2; blue, 27 }  ,fill opacity=1 ] (151,138) -- (171,138) -- (171,158) -- (151,158) -- cycle ;
    \draw  [fill={rgb, 255:red, 208; green, 2; blue, 27 }  ,fill opacity=1 ] (284,177) -- (304,177) -- (304,197) -- (284,197) -- cycle ;
    \draw  [fill={rgb, 255:red, 208; green, 2; blue, 27 }  ,fill opacity=1 ] (244,97) -- (264,97) -- (264,117) -- (244,117) -- cycle ;
    \draw  [fill={rgb, 255:red, 208; green, 2; blue, 27 }  ,fill opacity=1 ] (244,177) -- (264,177) -- (264,197) -- (244,197) -- cycle ;
    \draw  [fill={rgb, 255:red, 208; green, 2; blue, 27 }  ,fill opacity=1 ] (324,177) -- (344,177) -- (344,197) -- (324,197) -- cycle ;
    \draw  [fill={rgb, 255:red, 208; green, 2; blue, 27 }  ,fill opacity=1 ] (324,97) -- (344,97) -- (344,117) -- (324,117) -- cycle ;
    \draw  [fill={rgb, 255:red, 208; green, 2; blue, 27 }  ,fill opacity=1 ] (371,177) -- (391,177) -- (391,197) -- (371,197) -- cycle ;
    \draw  [fill={rgb, 255:red, 208; green, 2; blue, 27 }  ,fill opacity=1 ] (371,137) -- (391,137) -- (391,157) -- (371,157) -- cycle ;
    \draw  [fill={rgb, 255:red, 208; green, 2; blue, 27 }  ,fill opacity=1 ] (411,177) -- (431,177) -- (431,197) -- (411,197) -- cycle ;
    \draw  [fill={rgb, 255:red, 208; green, 2; blue, 27 }  ,fill opacity=1 ] (431,177) -- (451,177) -- (451,197) -- (431,197) -- cycle ;
    \draw  [fill={rgb, 255:red, 208; green, 2; blue, 27 }  ,fill opacity=1 ] (451,137) -- (471,137) -- (471,157) -- (451,157) -- cycle ;
    \draw  [fill={rgb, 255:red, 208; green, 2; blue, 27 }  ,fill opacity=1 ] (431,117) -- (451,117) -- (451,137) -- (431,137) -- cycle ;
    \draw  [fill={rgb, 255:red, 208; green, 2; blue, 27 }  ,fill opacity=1 ] (391,97) -- (411,97) -- (411,117) -- (391,117) -- cycle ;
    \draw  [fill={rgb, 255:red, 208; green, 2; blue, 27 }  ,fill opacity=1 ] (371,97) -- (391,97) -- (391,117) -- (371,117) -- cycle ;
    \draw  [fill={rgb, 255:red, 208; green, 2; blue, 27 }  ,fill opacity=1 ] (451,117.5) -- (471,117.5) -- (471,137.5) -- (451,137.5) -- cycle ;
    \draw    (275,127.5) -- (255.43,108.4) ;
    \draw [shift={(254,107)}, rotate = 44.31] [color={rgb, 255:red, 0; green, 0; blue, 0 }  ][line width=0.75]    (10.93,-3.29) .. controls (6.95,-1.4) and (3.31,-0.3) .. (0,0) .. controls (3.31,0.3) and (6.95,1.4) .. (10.93,3.29)   ;
    \draw    (294,168.5) -- (294,185) ;
    \draw [shift={(294,187)}, rotate = 270] [color={rgb, 255:red, 0; green, 0; blue, 0 }  ][line width=0.75]    (10.93,-3.29) .. controls (6.95,-1.4) and (3.31,-0.3) .. (0,0) .. controls (3.31,0.3) and (6.95,1.4) .. (10.93,3.29)   ;
    \draw    (315.5,147.75) -- (332,147.08) ;
    \draw [shift={(334,147)}, rotate = 177.68] [color={rgb, 255:red, 0; green, 0; blue, 0 }  ][line width=0.75]    (10.93,-3.29) .. controls (6.95,-1.4) and (3.31,-0.3) .. (0,0) .. controls (3.31,0.3) and (6.95,1.4) .. (10.93,3.29)   ;
    \draw    (316,168) -- (332.62,185.55) ;
    \draw [shift={(334,187)}, rotate = 226.55] [color={rgb, 255:red, 0; green, 0; blue, 0 }  ][line width=0.75]    (10.93,-3.29) .. controls (6.95,-1.4) and (3.31,-0.3) .. (0,0) .. controls (3.31,0.3) and (6.95,1.4) .. (10.93,3.29)   ;
    \draw    (275,167.5) -- (255.47,185.64) ;
    \draw [shift={(254,187)}, rotate = 317.12] [color={rgb, 255:red, 0; green, 0; blue, 0 }  ][line width=0.75]    (10.93,-3.29) .. controls (6.95,-1.4) and (3.31,-0.3) .. (0,0) .. controls (3.31,0.3) and (6.95,1.4) .. (10.93,3.29)   ;
    \draw    (315,127.5) -- (332.64,108.47) ;
    \draw [shift={(334,107)}, rotate = 132.83] [color={rgb, 255:red, 0; green, 0; blue, 0 }  ][line width=0.75]    (10.93,-3.29) .. controls (6.95,-1.4) and (3.31,-0.3) .. (0,0) .. controls (3.31,0.3) and (6.95,1.4) .. (10.93,3.29)   ;
    \draw    (274.5,147.25) -- (256,147.02) ;
    \draw [shift={(254,147)}, rotate = 0.7] [color={rgb, 255:red, 0; green, 0; blue, 0 }  ][line width=0.75]    (10.93,-3.29) .. controls (6.95,-1.4) and (3.31,-0.3) .. (0,0) .. controls (3.31,0.3) and (6.95,1.4) .. (10.93,3.29)   ;
    \draw    (294.5,127.25) -- (294.05,109) ;
    \draw [shift={(294,107)}, rotate = 88.59] [color={rgb, 255:red, 0; green, 0; blue, 0 }  ][line width=0.75]    (10.93,-3.29) .. controls (6.95,-1.4) and (3.31,-0.3) .. (0,0) .. controls (3.31,0.3) and (6.95,1.4) .. (10.93,3.29)   ;
    \draw    (441,127) -- (459,127.45) ;
    \draw [shift={(461,127.5)}, rotate = 181.43] [color={rgb, 255:red, 0; green, 0; blue, 0 }  ][line width=0.75]    (10.93,-3.29) .. controls (6.95,-1.4) and (3.31,-0.3) .. (0,0) .. controls (3.31,0.3) and (6.95,1.4) .. (10.93,3.29)   ;
    \draw    (402,167) -- (382.45,185.62) ;
    \draw [shift={(381,187)}, rotate = 316.4] [color={rgb, 255:red, 0; green, 0; blue, 0 }  ][line width=0.75]    (10.93,-3.29) .. controls (6.95,-1.4) and (3.31,-0.3) .. (0,0) .. controls (3.31,0.3) and (6.95,1.4) .. (10.93,3.29)   ;
    \draw    (421,147) -- (383,147) ;
    \draw [shift={(381,147)}, rotate = 360] [color={rgb, 255:red, 0; green, 0; blue, 0 }  ][line width=0.75]    (10.93,-3.29) .. controls (6.95,-1.4) and (3.31,-0.3) .. (0,0) .. controls (3.31,0.3) and (6.95,1.4) .. (10.93,3.29)   ;
    \draw    (401,147) -- (381.89,108.79) ;
    \draw [shift={(381,107)}, rotate = 63.43] [color={rgb, 255:red, 0; green, 0; blue, 0 }  ][line width=0.75]    (10.93,-3.29) .. controls (6.95,-1.4) and (3.31,-0.3) .. (0,0) .. controls (3.31,0.3) and (6.95,1.4) .. (10.93,3.29)   ;
    \draw    (421,167) -- (421,185) ;
    \draw [shift={(421,187)}, rotate = 270] [color={rgb, 255:red, 0; green, 0; blue, 0 }  ][line width=0.75]    (10.93,-3.29) .. controls (6.95,-1.4) and (3.31,-0.3) .. (0,0) .. controls (3.31,0.3) and (6.95,1.4) .. (10.93,3.29)   ;
    \draw    (440,167) -- (440.9,185) ;
    \draw [shift={(441,187)}, rotate = 267.14] [color={rgb, 255:red, 0; green, 0; blue, 0 }  ][line width=0.75]    (10.93,-3.29) .. controls (6.95,-1.4) and (3.31,-0.3) .. (0,0) .. controls (3.31,0.3) and (6.95,1.4) .. (10.93,3.29)   ;
    \draw    (441,146) -- (459,146.9) ;
    \draw [shift={(461,147)}, rotate = 182.86] [color={rgb, 255:red, 0; green, 0; blue, 0 }  ][line width=0.75]    (10.93,-3.29) .. controls (6.95,-1.4) and (3.31,-0.3) .. (0,0) .. controls (3.31,0.3) and (6.95,1.4) .. (10.93,3.29)   ;
    \draw    (401,127) -- (401,109) ;
    \draw [shift={(401,107)}, rotate = 90] [color={rgb, 255:red, 0; green, 0; blue, 0 }  ][line width=0.75]    (10.93,-3.29) .. controls (6.95,-1.4) and (3.31,-0.3) .. (0,0) .. controls (3.31,0.3) and (6.95,1.4) .. (10.93,3.29)   ;
    \draw    (421,127) -- (439,127) ;
    \draw [shift={(441,127)}, rotate = 180] [color={rgb, 255:red, 0; green, 0; blue, 0 }  ][line width=0.75]    (10.93,-3.29) .. controls (6.95,-1.4) and (3.31,-0.3) .. (0,0) .. controls (3.31,0.3) and (6.95,1.4) .. (10.93,3.29)   ;
    \end{tikzpicture}
    }
    \caption{Illustration (inspired by \cite{dcn}) of sampling points in standard, dilated and deformable convolution with kernel size $3\times3$.}
    \label{fig:convs}
\end{figure}

We propose and test the use of deformable convolutions.
Originally presented in \cite{dcn}, this CNN layer has been used in several computer vision problems but, to the best of our knowledge, we are the first ones to apply it for radio map prediction.
Similarly to a dilated convolution, it allows to enlarge the receptive field by sampling the input at positions further away. 
The sampling points are not fixed  (Fig. \ref{fig:convs}), instead the offset compared to a standard convolution is computed from the input with learnable parameters.
Intuitively, this should make it easier to propagate information in arbitrary directions compared to dilated convolutions.
We use this layer as a replacement for some of the convolutions in an otherwise fairly standard UNet and, following \cite{dcn}, call the resulting model UNetDCN.

In Table \ref{table:complexity} we compare the complexity of the different architectures in different aspects. 
RadioUNet is the most light-weight and the fastest model. 
PMNet has the highest complexity in terms of multiply-accumulate operations (MACs) but a lower number of parameters compared to UNetDCN.
We have noticed that the training of UNetDCN is the slowest, but at inference it is faster than PMNet.
We suspect that the implementation of deformable convolutions is less efficient than the highly optimized implementations of standard layers on modern GPUs.

\begin{table*}
    \centering
    \begin{tabular}{|c|c|c|c|}
        \hline
        Model                   &   RadioUNet \cite{radiounet}    &   PMNet   \cite{pmnet}      &   UNetDCN    \\
        \hline
        \hline
        \# parameters$^{a}$     &  $10.93$M                         &   $33.4$M                     &  $69.8$M \\
        \# MACs$^{a}$           &  $7.7$G                           &   $50.7$G                     &  $14.7$G \\
        inference time per sample& $0.0095$s                     &   $0.0204$s                   &  $0.0161$s \\
        \hline
    \end{tabular}
    \caption{Complexity of the considered architectures for the default inputs and batch size 1\\$^{a}$calculated using DeepSpeed \cite{deepspeed}.}
    \label{table:complexity}
\end{table*}

\subsection{INPUT FEATURES}\label{sec:input_features}
\begin{figure*}[h]
    \centering
    \begin{subfigure}[t]{0.13\linewidth}
        \includegraphics[width=\linewidth]{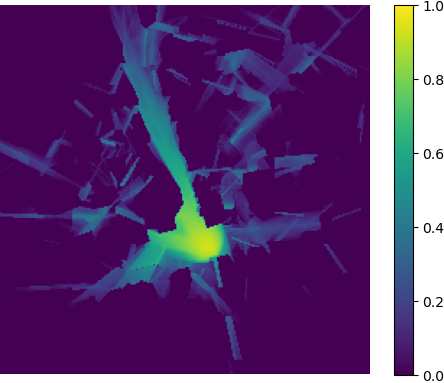}
        \caption{}
        \label{fig:input_features:target}
    \end{subfigure}
    \begin{subfigure}[t]{0.13\linewidth}
        \includegraphics[width=\linewidth]{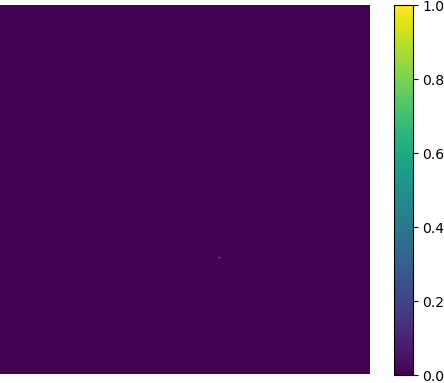}
        \caption{}
        \label{fig:input_features:tx_one_hot}
    \end{subfigure}
    \begin{subfigure}[t]{0.13\linewidth}
        \includegraphics[width=\linewidth]{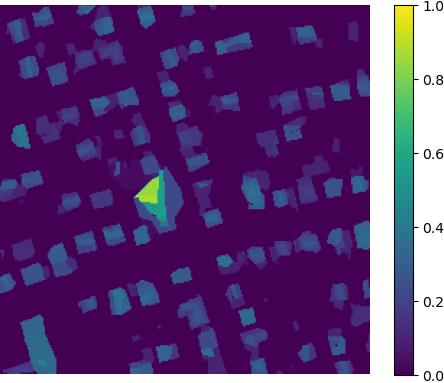}
        \caption{}
        \label{fig:input_features:build_ndsm}
    \end{subfigure}
    \begin{subfigure}[t]{0.13\linewidth}
        \includegraphics[width=\linewidth]{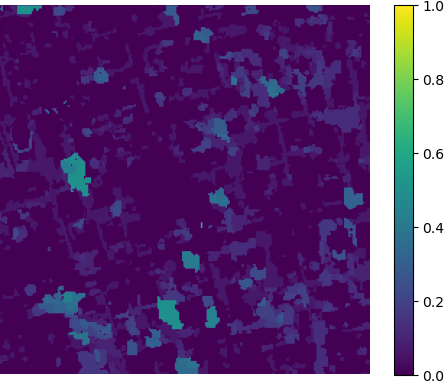}
        \caption{}
        \label{fig:input_features:veg_ndsm}
    \end{subfigure}
    \begin{subfigure}[t]{0.13\linewidth}
        \includegraphics[width=\linewidth]{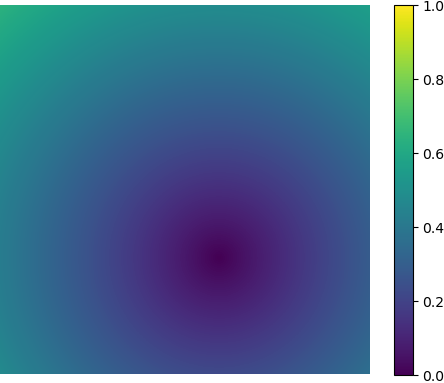}
        \caption{}
        \label{fig:input_features:dist2d}
    \end{subfigure}
    \begin{subfigure}[t]{0.135\linewidth}
        \includegraphics[width=\linewidth]{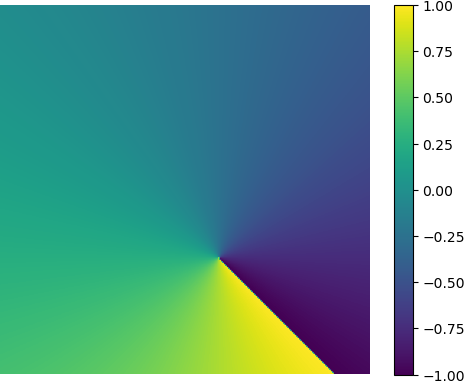}
        \caption{}
        \label{fig:input_features:azimuth}
    \end{subfigure}
    \begin{subfigure}[t]{0.13\linewidth}
        \includegraphics[width=\linewidth]{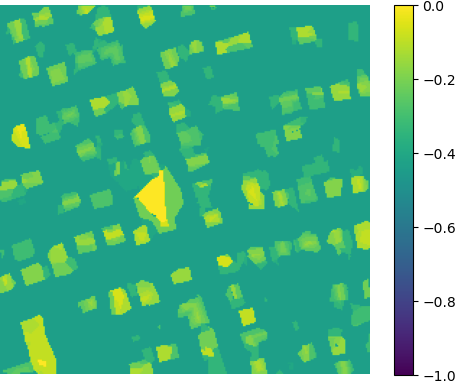}
        \caption{}
        \label{fig:input_features:build_rel}
    \end{subfigure}
    \begin{subfigure}[t]{0.13\linewidth}
        \includegraphics[width=\linewidth]{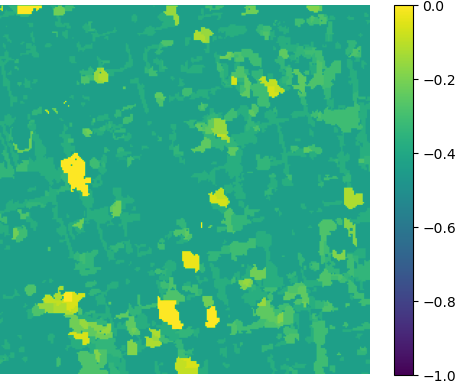}
        \caption{}
        \label{fig:input_features:veg_rel}
    \end{subfigure}
    \begin{subfigure}[t]{0.13\linewidth}
        \includegraphics[width=\linewidth]{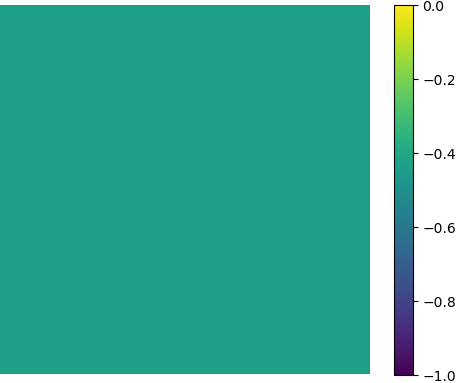}
        \caption{}
        \label{fig:input_features:floor_rel}
    \end{subfigure}
    \begin{subfigure}[t]{0.13\linewidth}
        \includegraphics[width=\linewidth]{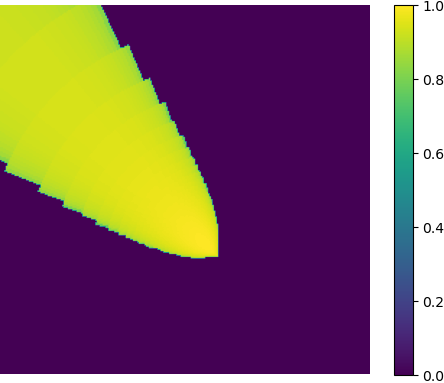}
        \caption{}
        \label{fig:input_features:gain_floor}
    \end{subfigure}
    \begin{subfigure}[t]{0.13\linewidth}
        \includegraphics[width=\linewidth]{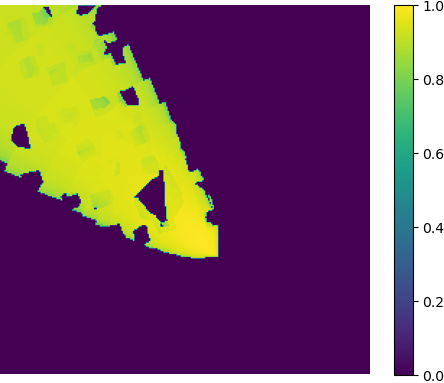}
        \caption{}
        \label{fig:input_features:gain_top}
    \end{subfigure}
    \begin{subfigure}[t]{0.13\linewidth}
        \includegraphics[width=\linewidth]{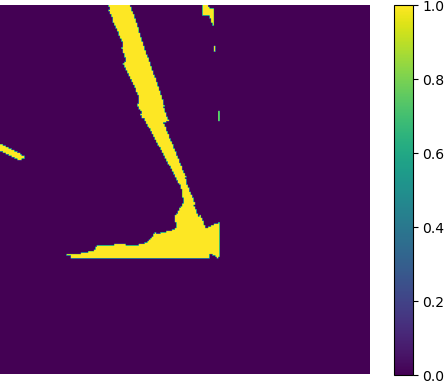}
        \caption{}
        \label{fig:input_features:los_floor}
    \end{subfigure}
    \begin{subfigure}[t]{0.13\linewidth}
        \includegraphics[width=\linewidth]{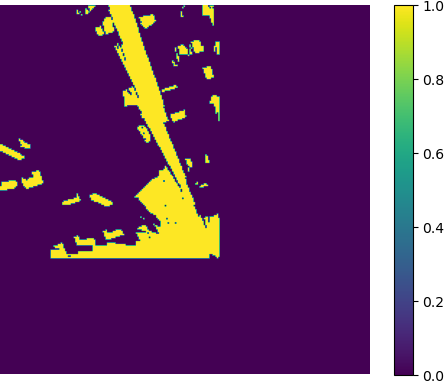}
        \caption{}
        \label{fig:input_features:los_top}
    \end{subfigure}
    \begin{subfigure}[t]{0.135\linewidth}
        \includegraphics[width=\linewidth]{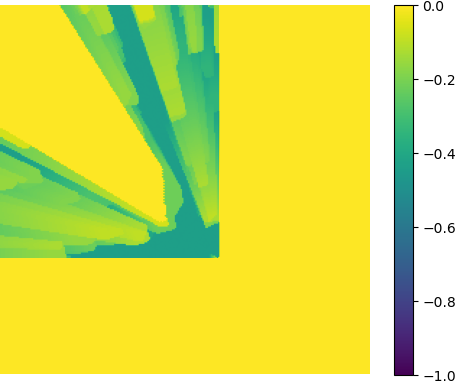}
        \caption{}
        \label{fig:input_features:los_min}
    \end{subfigure}
    \caption{Example radio map \subref{fig:input_features:target} and (normalized) input features. 
    Baseline Tx and city geometry encoding: \subref{fig:input_features:tx_one_hot} Tx location, \subref{fig:input_features:build_ndsm} building nDSM, \subref{fig:input_features:veg_ndsm} vegetation nDSM. 
    Cylindrical coordinates: \subref{fig:input_features:dist2d} 2D distance to Tx, \subref{fig:input_features:azimuth} azimuth angle, \subref{fig:input_features:build_rel} building height relative to Tx, \subref{fig:input_features:veg_rel} vegetation height relative to Tx, \subref{fig:input_features:floor_rel} ground height relative to Tx.
        Tx antenna pattern: \subref{fig:input_features:gain_floor} pattern projected onto the ground, \subref{fig:input_features:gain_top} pattern projected onto building top.
        Line-of-sight (LoS) information: \subref{fig:input_features:los_floor} binary LoS ground, \subref{fig:input_features:los_top} binary LoS building top, \subref{fig:input_features:los_min} ours (for relative heights from cylindrical coordinates).}
    \label{fig:input_features}
\end{figure*}

Following the related literature (Section \ref{sec:related_works}), we aim to encode all relevant parameters that change between the different simulations in 2D images, where each pixel corresponds to a location on the map.
In the following, we provide an explanation of the input features we consider.
Several examples for a sample from the dataset are shown in Fig. \ref{fig:input_features} and further details of the implementation can be found in the code.
All inputs are normalized to values in $[-1, 1]$ before being fed to the CNN.

\subsubsection{ENVIRONMENT ENCODING AND TX LOCATION}\label{sec:geometry}
In the $2D$ setting, the positions of the Tx, buildings and potentially other objects are often given in a \textit{one-hot encoding} in separate binary tensors, where a $1$ indicates presence of the Tx or object and $0$ absence in the location corresponding to each pixel (see e.g. \cite{radiounet}).
Since in our work the height of the Tx, buildings and vegetation are relevant, we assign these as the values to the pixels instead, as it is done in \cite{radiotrans} or \cite{fadenet} for example (Fig. \ref{fig:input_features}\subref{fig:input_features:tx_one_hot}-\subref{fig:input_features:veg_ndsm}).
The sparsity of the tensor representing the Tx location is potentially problematic, as the standard layers used in CNNs inherently have a very limited field of view and it therefore takes several convolutional and down-sampling operations to spread the information to other parts of the map.

The authors of \cite{radiotrans} propose to tackle this issue by making the information about the spatial position of each pixel together with the location of the Tx explicitly available to the model, in the form of constant input tensors for the $x, y$ and $z$ coordinates of the Tx and two tensors showing the $x$ and $y$ coordinates of each pixel.
Although they state that their main intention was to find an alternative to the usual positional embedding in vision transformer layers, we have found that this approach also improves the performance of other CNN models.
They denote their idea as \textit{Grid Anchor (GA)}.
Inspired by this approach, we have tested alternatives to encode the positions and heights of each pixel and object \textit{relative to the Tx and the Tx orientation}.

First, we test an encoding in \textit{cylindrical coordinates} (Fig. \ref{fig:input_features}\subref{fig:input_features:dist2d}-\subref{fig:input_features:floor_rel}).
Intuitively, having access to the azimuth angles from the Tx to each point on the map makes it easier to determine whether an object interferes the line of sight path to an Rx or another object, since two locations lie on the same straight line from the Tx if and only if the corresponding azimuth angles are (approximately) equal.
The orientation is chosen so that the direction of the main lobe of the Tx always corresponds to the azimuth angle $0$, to help the network identify it.
The distance in the $x-y$ plane included in the cylindrical coordinate system is expected to facilitate the calculation of the free space path loss along line of sight paths.
This last input is also proposed in \cite{qiu22} in two variations, but the authors do not compare to the performance without any distance information.

Similarly, we have expressed the city map in an \textit{euclidian coordinate system} with origin in the Tx location.
More precisely, the planar coordinates of each pixel or tensor element are given relative to the Tx position as a positive or negative number, indicating in which direction and how far the Tx is located.
In the same manner, we encode the height of the ground, buildings and vegetation relative to the Tx height. 
This is very closely related to GA and can be derived from there with a few easy calculations. 

Lastly, we also train networks with a \textit{spherical coordinate system} defined in an analogous way. 
Note that already the authors of \cite{plnet} have proposed to provide the network with the azimuth and elevation angle of each pixel at ground height relative to the Tx location and main lobe, with the intention to encode the antenna orientation this way.
In our extension, we define elevation angle and 3D distance for ground, buildings and vegetation in each point on the map and azimuth angle for each spatial location to obtain a complete description in spherical coordinates.
Intuitively, this makes it straight forward to decide if an object can block or attenuate the direct path to another object or point on the ground by simply comparing the elevation angles (see Fig. \ref{fig:los-min}).
Furthermore, the authors of \cite{plnet} do not perform any experiments without these inputs for comparison, so it is not clear how effective they are.

All of these options in principle provide the same information about the environment but in different representations, that can be more or less suitable for the given task.

\subsubsection{TX ANTENNA PATTERN AND ORIENTATION}\label{sec:antennas}
To link the antenna pattern to the spatial positions, we use the spherical coordinate system described above and look up the gain in $dB$ corresponding to azimuth and tilt angle for each point on the ground, as it is done in \cite{plnet} as well.
In this way, we are potentially losing important information about the gain in the same spatial position but at a higher elevation, which might be relevant for predicting the strength of a reflected path.

In a first attempt to overcome this problem, we have defined an input tensor additionally containing the gain corresponding to the highest point of the building in each position (Fig. \ref{fig:input_features}\subref{fig:input_features:gain_floor},\subref{fig:input_features:gain_top}).
To provide more detailed information about the radiation pattern, we also test encoding the antenna gain in \textit{slices}.
For this, we project the antenna pattern onto several planes parallel to the ground at e.g. $0, 4,\ldots, 28$m height instead.
The idea of encoding 3D information in this way has been used in \cite{slices} as well to express a 3D indoor scenario and generate radio maps at different heights.
We have also run some tests on encoding the height of the objects on our maps in that way but no significant improvement was observed.

Furthermore, we have tested providing the model with an estimate of the free space path loss in dB (later called \textit{FSPL}) instead of the antenna gain, pointwise defined as 
\begin{equation}
    \FSPL    =   g - 20 \cdot\log_{10}(d),
\end{equation}
where $g$ is the antenna gain in dB and $d$ the distance in m (as in Fig. \ref{fig:input_features:dist2d}).

\subsubsection{LINE-OF-SIGHT}\label{sec:los}
Since the strongest signal is often received in areas with LoS to the Tx and the direct paths mostly dominate other paths undergoing reflections or diffractions, recognizing whether the direct line from the Tx to a point is obstructed is one of the key aspects to predict an accurate radio map.
Some authors (see e.g. \cite{fadenet}) have therefore suggested to aid the models by providing a binary encoded LoS map, where the value in each spatial position is $1$ if there is a direct LoS path from the Tx to the location on ground level and $0$ otherwise.

\begin{figure}[!ht]
\resizebox{\linewidth}{!}{
\tikzset{every picture/.style={line width=0.75pt}} 
\begin{tikzpicture}[x=0.75pt,y=0.75pt,yscale=-1,xscale=1]

\draw  [fill={rgb, 255:red, 228; green, 8; blue, 8 }  ,fill opacity=1 ] (459,83.5) -- (529,83.5) -- (529,250) -- (459,250) -- cycle ;
\draw    (50,250) -- (553.33,250) ;
\draw  [fill={rgb, 255:red, 55; green, 255; blue, 0 }  ,fill opacity=1 ] (461,54) -- (482.5,54) -- (482.5,77) -- (461,77) -- cycle ;
\draw  [fill={rgb, 255:red, 228; green, 8; blue, 8 }  ,fill opacity=1 ] (112.5,139) -- (238,139) -- (238,250) -- (112.5,250) -- cycle ;
\draw  [fill={rgb, 255:red, 228; green, 8; blue, 8 }  ,fill opacity=1 ] (301.5,175) -- (400,175) -- (400,250) -- (301.5,250) -- cycle ;
\draw    (461.5,68) -- (237,216.75) ;
\draw    (482.5,77) -- (489.5,83.75) ;
\draw    (461,77) -- (459,83.5) ;
\draw    (238,139) -- (461.5,68) ;
\draw [color={rgb, 255:red, 74; green, 144; blue, 226 }  ,draw opacity=1 ][fill={rgb, 255:red, 0; green, 76; blue, 255 }  ,fill opacity=1 ][line width=1.5]    (260,215.88) -- (260,249.63) ;
\draw [color={rgb, 255:red, 74; green, 144; blue, 226 }  ,draw opacity=1 ][fill={rgb, 255:red, 0; green, 76; blue, 255 }  ,fill opacity=1 ][line width=1.5]    (251,215.75) -- (270,215.5) ;
\draw [color={rgb, 255:red, 74; green, 144; blue, 226 }  ,draw opacity=1 ][line width=1.5]    (90.5,140) -- (90,250) ;
\draw [color={rgb, 255:red, 74; green, 144; blue, 226 }  ,draw opacity=1 ][fill={rgb, 255:red, 0; green, 76; blue, 255 }  ,fill opacity=1 ][line width=1.5]    (250.5,249.75) -- (270,250) ;
\draw [color={rgb, 255:red, 74; green, 144; blue, 226 }  ,draw opacity=1 ][fill={rgb, 255:red, 0; green, 76; blue, 255 }  ,fill opacity=1 ][line width=1.5]    (81,139.75) -- (100,140) ;
\draw [color={rgb, 255:red, 74; green, 144; blue, 226 }  ,draw opacity=1 ][fill={rgb, 255:red, 0; green, 76; blue, 255 }  ,fill opacity=1 ][line width=1.5]    (80.5,249.75) -- (100,250) ;
\draw  [fill={rgb, 255:red, 248; green, 231; blue, 28 }  ,fill opacity=1 ] (234.38,250.25) .. controls (234.25,248.25) and (235.75,246.52) .. (237.75,246.38) .. controls (239.75,246.25) and (241.48,247.75) .. (241.62,249.75) .. controls (241.75,251.75) and (240.25,253.48) .. (238.25,253.62) .. controls (236.25,253.75) and (234.52,252.25) .. (234.38,250.25) -- cycle ;
\draw [color={rgb, 255:red, 245; green, 166; blue, 35 }  ,draw opacity=1 ]   (461.07,67.28) -- (348.11,87.79) ;
\draw [shift={(346.14,88.14)}, rotate = 349.71] [color={rgb, 255:red, 245; green, 166; blue, 35 }  ,draw opacity=1 ][line width=0.75]    (10.93,-3.29) .. controls (6.95,-1.4) and (3.31,-0.3) .. (0,0) .. controls (3.31,0.3) and (6.95,1.4) .. (10.93,3.29)   ;
\draw  [fill={rgb, 255:red, 245; green, 166; blue, 35 }  ,fill opacity=1 ] (449.49,75.24) .. controls (449.49,75.24) and (449.49,75.24) .. (449.49,75.24) .. controls (441.46,79.89) and (428.77,81.88) .. (421.14,79.68) .. controls (413.52,77.47) and (413.85,71.92) .. (421.88,67.27) .. controls (429.91,62.62) and (442.6,60.63) .. (450.22,62.83) .. controls (458.46,65.21) and (462.08,66.7) .. (461.07,67.28) .. controls (462.02,67.56) and (458.16,70.22) .. (449.49,75.24) -- cycle ;
\draw  [draw opacity=0][line width=1.5]  (396.21,110.88) .. controls (395.21,109.38) and (394.25,107.83) .. (393.34,106.24) .. controls (388.65,98.09) and (385.5,89.49) .. (383.79,80.78) -- (461.07,67.28) -- cycle ; \draw  [color={rgb, 255:red, 144; green, 19; blue, 254 }  ,draw opacity=1 ][line width=1.5]  (396.21,110.88) .. controls (395.21,109.38) and (394.25,107.83) .. (393.34,106.24) .. controls (388.65,98.09) and (385.5,89.49) .. (383.79,80.78) ;  
\draw  [draw opacity=0][line width=1.5]  (391.86,89.65) .. controls (390.67,86.42) and (389.68,83.16) .. (388.89,79.89) -- (461.07,67.28) -- cycle ; \draw  [color={rgb, 255:red, 144; green, 19; blue, 254 }  ,draw opacity=1 ][line width=1.5]  (391.86,89.65) .. controls (390.67,86.42) and (389.68,83.16) .. (388.89,79.89) ;  

\draw (262.01,223.38) node [anchor=north west][inner sep=0.75pt]  [font=\large,rotate=-0.23,xslant=-0.01] [align=left] {\textcolor[rgb]{0.29,0.56,0.89}{min}};
\draw (74.47,174.93) node [anchor=north west][inner sep=0.75pt]  [font=\large,rotate=-0.23,xslant=-0.01] [align=left] {\textcolor[rgb]{0.29,0.56,0.89}{z}};
\end{tikzpicture}

}
    
\caption{Illustration of our LoS encoding and elevation angles -- 
vertical cut through a simple city environment with buildings in red, 
Tx in green with antenna pattern and main direction in orange, 
location of interest in yellow, 
height of the building at this location (z) and smallest height value for which LoS is not obstructed (min) in blue, 
elevation angles in purple (corresponding to the building in the location of interest and the left corner of the building in the middle).}
\label{fig:los-min}
\end{figure}
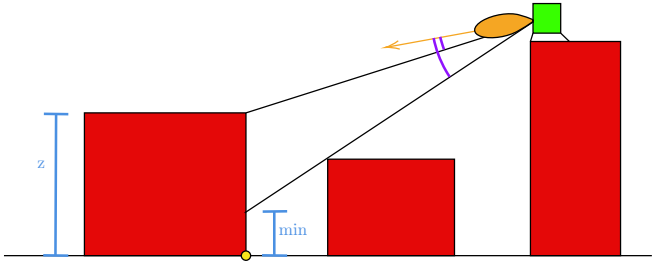

Similar to the antenna encoding, we have first extended this by additionally providing the information whether the top of the building in each pixel is visible from the Tx or not (\textit{binary} in Table \ref{table:geometry_los}). 
Intuitively, this may help the network to identify reflecting walls and diffractions from the edges of buildings towards the ground.
Still, also this method is not suitable to encode the 3D propagation of the direct paths from the Tx completely, since it does not provide information on, for example, the height at which a wall may reflect a path.
As a solution, we consider providing the models with a tensor showing the smallest height level that is visible from the Tx in each position (denoted as \textit{ours} in Table \ref{table:geometry_los} and Fig. \ref{fig:input_features:los_min}, see Fig. \ref{fig:los-min} for a graphical explanation).
For the relative coordinate systems, this value is adjusted to the minimum height relative to the height of the Tx, respectively the maximum elevation angle in spherical coordinates instead, see Fig. \ref{fig:input_features}\subref{fig:input_features:los_floor}-\subref{fig:input_features:los_min} for an example. 
Note that our approach to calculate LoS does not take the full antenna pattern into account, but locations outside of the cone defined by the first null beam width are considered to be in non-LoS.

\section{NUMERICAL EXPERIMENTS}\label{sec:numerical_results}
We use about $80\%$ of the samples as the training set, $10\%$ for validation and $10\%$ for testing, making sure that the city maps do not overlap between the different sets.
The validation set is applied to first reduce the learning rate and later stop training when the loss stagnates, and to determine the best model weights to be saved.
The test set is only used at the very end to evaluate the final performance and generalization capability.
During training, we apply random flips and rotations as data augmentations.

All models are trained with respect to mean-squared error (MSE).
We measure the final performance in terms of root-mean-square error (RMSE) in grayscale.
Recall that the grayscale radio map values in $[0,1]$ stem from an affine transformation of the possible path loss values in dB, which lie in $[-127\text{dB}, -50\text{dB}]$, hence the RMSE in dB can be obtained from multiplying the RMSE in grayscale by a factor of $77$.
Furthermore, we report the normalized mean-squared error (NMSE) after conversion to dB scale, in order to emphasize the importance of samples with a high signal power (see \cite{radiounet}).

Our code is implemented in PyTorch Lightning. 
The models are trained with a batch size of $32$ and Adam optimizer with an initial learning rate of $10^{-4}$.
Typically, fitting converges within the first $40$ epochs.
All models are trained on A100 GPUs but otherwise differing hardware, as we work on a cluster with different nodes.
Reported losses are generated on the test set.
Learning rate and scheduling had been found to work well in preliminary experiments with all models and a few input combinations.
However, after the final training runs we have noticed that UNetDCN did not converge properly in three of the considered scenarios, for these cases we have restarted training with lower learning rate and scheduling adjusted accordingly.
We have also verified once more that the other models do not profit from the same modification.
\begin{table*}[!htb]
    \centering
    \begin{tabular}{|c|c||c|c|c|c|c|c|}
        \hline
        \multicolumn{2}{|c|}{Model} &   \multicolumn{2}{c|}{RadioUNet \cite{radiounet}}&   \multicolumn{2}{c|}{PMNet \cite{pmnet}}     &   \multicolumn{2}{c|}{UNetDCN}\\    
        \hline
        \cline{1-2}
        Geometry                & LoS       & RMSE              & NMSE                &   RMSE                &   NMSE                  &     RMSE              &     NMSE\\
        \hline
        \multirow{3}{*}{Basic}  &   -       &   $0.0713$        &   $0.00217$         &   ${0.0658}$          &   ${0.00184}$           &   $\textbf{0.0645}$   &   $\textbf{0.00178}$     \\
                                &   binary  &   $0.0700$        &   ${0.00208}$       &   $0.0654$            &   $0.00181$             &   $\textbf{0.0635}$   &   $\textbf{0.00171}$     \\
                                &   ours     &   $0.0686$        &   $0.00201$         &   ${0.0635}$          &   $0.00171$             &   $\textbf{0.0621}$   &   $\textbf{0.00164}$ \\
        \hline
        \multirow{3}{*}{GA \cite{radiotrans}}     &   -       &   $0.0699$        &   $0.00208$         &   $0.0635$            &   $0.00171$    &   $\textbf{0.0631}$   &   $\textbf{0.00169}$  \\
                                &   binary  &   $0.0694$        &   $0.00204$         &   $\textbf{0.0623}$   &   $\textbf{0.00165}$    &   $0.0631$            &   $0.00169$\\
                                &   ours     &   $0.0682$        &   $0.00198$         &   $\textbf{0.0620}$   &   $\textbf{0.00163}$    &   {$0.0629$}   &   {$0.00167$}  \\
        \hline
        \multirow{3}{*}{Euclidian coordinates}  &   -   &   $\textit{0.0697}$&  ${0.00207}$&   ${0.0632}$   &   ${0.00169}$    &   $\textbf{\textit{0.0623}}$   &   $\textbf{\textit{0.00164}}$ \\
                                &   binary  &   $0.0696$        &   $0.00206$         &   $\textbf{0.0625}$   &   $\textbf{0.00165}$    &   $\textbf{0.0625}$   &   $0.00166$ \\
                                &   ours     &   $\underline{0.0681}$&$\underline{0.00197}$&   $0.0623$        &   ${{0.00165}}$         &   $\textbf{0.0618}$   &   $\textbf{0.00162}$    \\
        \hline
        \multirow{3}{*}{Cylindrical coordinates}   &   -&   $\textit{0.0697}$&$\textit{0.00206}$        &   $\textbf{\textit{0.0630}}$   &   $\textbf{\textit{0.00168}}$             &  $0.0633$             &   $0.00170$  \\
                                &   binary  &   $0.0698$        &   $0.00207$         &   $\textbf{0.0626}$   &   $\textbf{0.00166}$    &  {$0.0634$}    &   {$0.00170$}  \\
                                &   ours     &   $0.0683$        &   $0.00198$         &   $\underline{0.0615}$&   $\underline{0.00160}$ &  $\underline{\textbf{0.0611}}$&   $\underline{0.00158}$     \\
        \hline
        \multirow{3}{*}{Spherical coordinates}   &   -  &   $0.0716$        &   $0.00218$         &   $0.0641$            &   $\textbf{0.00174}$    &   $\textbf{0.0640}$   &  $\textbf{0.00174}$      \\
                                &   binary  &   $0.0689$        &   $0.00202$         &   $\textbf{0.0638}$   &   $\textbf{0.00172}$    &   $0.0643$            &  $0.00175$       \\
                                &   ours    &   $0.0690$        &   $0.00202$         &   $0.0638$            &   ${0.00173}$           &   $\textbf{0.0626}$   &  $\textbf{0.00166}$  \\
        \hline
    \end{tabular}
    \caption{City geometry and LoS -- lowest errors in each column (input) without LoS information in \textit{italics}, with LoS information \underline{underlined} and per row (model) in \textbf{bold}.}
    \label{table:geometry_los}
\end{table*}

\begin{figure*}[!htb]
    \centering
    \begin{subfigure}[t]{0.12\textwidth}
    \includegraphics[width=\textwidth]{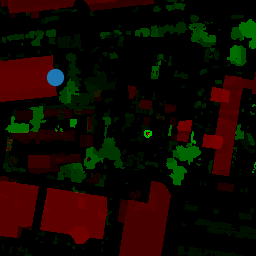}
    \end{subfigure}
    \begin{subfigure}[t]{0.12\textwidth}
    \includegraphics[width=\textwidth]{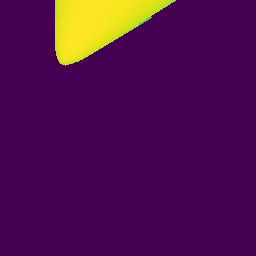}
    \end{subfigure}
    \begin{subfigure}[t]{0.033\textwidth}
        \includegraphics[width=\textwidth]{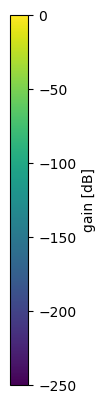}
    \end{subfigure}
    \begin{subfigure}[t]{0.12\textwidth}
    \includegraphics[width=\textwidth]{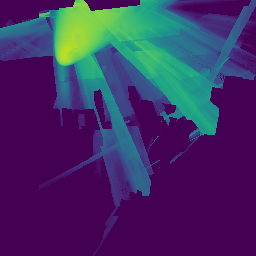}
    \end{subfigure}
    \begin{subfigure}[t]{0.12\textwidth}
    \includegraphics[width=\textwidth]{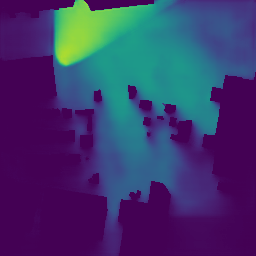}
    \caption*{0.0128}
    \end{subfigure}
    \begin{subfigure}[t]{0.12\textwidth}
    \includegraphics[width=\textwidth]{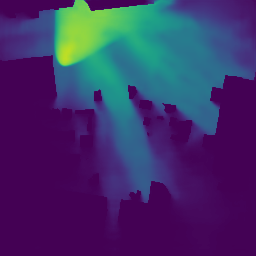}
    \caption*{0.0080}
    \end{subfigure}
    \begin{subfigure}[t]{0.12\textwidth}
    \includegraphics[width=\textwidth]{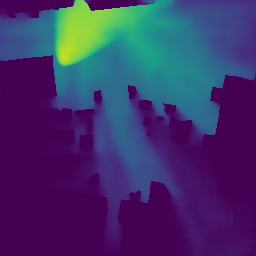}
    \caption*{0.0070}
    \end{subfigure}
    \begin{subfigure}[t]{0.033\textwidth}
        \includegraphics[width=\textwidth]{jaens5}
    \end{subfigure}
    \\
    \begin{subfigure}[t]{0.12\textwidth}
    \includegraphics[width=\textwidth]{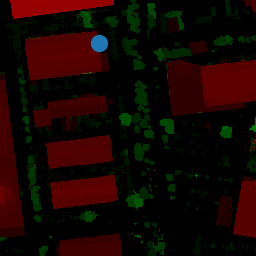}
    \end{subfigure}
    \begin{subfigure}[t]{0.12\textwidth}
    \includegraphics[width=\textwidth]{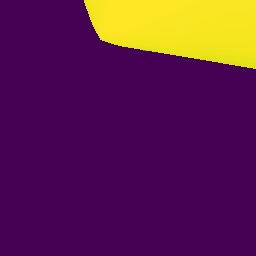}
    \end{subfigure}
    \begin{subfigure}[t]{0.033\textwidth}
        \includegraphics[width=\textwidth]{jaens25}
    \end{subfigure}
    \begin{subfigure}[t]{0.12\textwidth}
    \includegraphics[width=\textwidth]{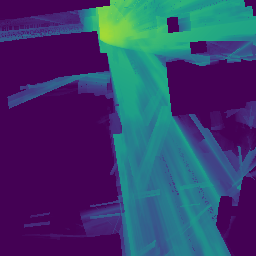}
    \end{subfigure}
    \begin{subfigure}[t]{0.12\textwidth}
    \includegraphics[width=\textwidth]{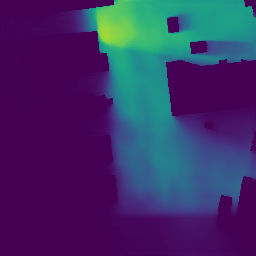}
    \caption*{0.0154}
    \end{subfigure}
    \begin{subfigure}[t]{0.12\textwidth}
    \includegraphics[width=\textwidth]{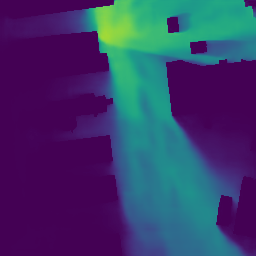}
    \caption*{0.0066}
    \end{subfigure}
    \begin{subfigure}[t]{0.12\textwidth}
    \includegraphics[width=\textwidth]{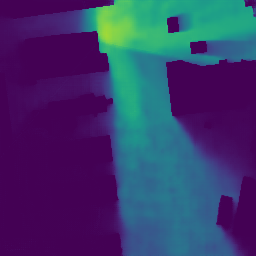}
    \caption*{0.0058}
    \end{subfigure}
    \begin{subfigure}[t]{0.033\textwidth}
        \includegraphics[width=\textwidth]{jaens5}
    \end{subfigure}
    \\
    \begin{subfigure}[t]{0.12\textwidth}
    \includegraphics[width=\textwidth]{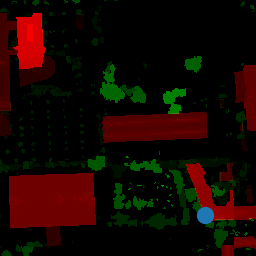}
    \end{subfigure}
    \begin{subfigure}[t]{0.12\textwidth}
    \includegraphics[width=\textwidth]{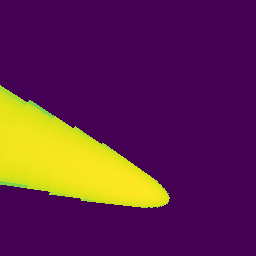}
    \end{subfigure}
    \begin{subfigure}[t]{0.033\textwidth}
        \includegraphics[width=\textwidth]{jaens25}
    \end{subfigure}
    \begin{subfigure}[t]{0.12\textwidth}
    \includegraphics[width=\textwidth]{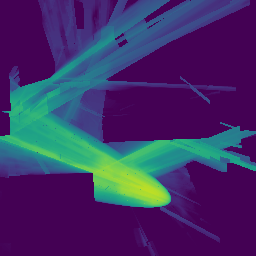}
    \end{subfigure}
    \begin{subfigure}[t]{0.12\textwidth}
    \includegraphics[width=\textwidth]{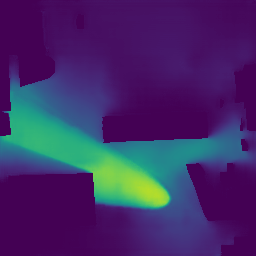}
    \caption*{0.0125}
    \end{subfigure}
    \begin{subfigure}[t]{0.12\textwidth}
    \includegraphics[width=\textwidth]{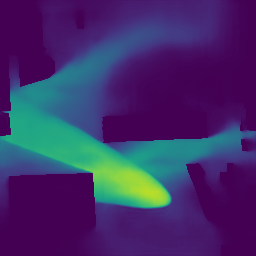}
    \caption*{0.0061}
    \end{subfigure}
    \begin{subfigure}[t]{0.12\textwidth}
    \includegraphics[width=\textwidth]{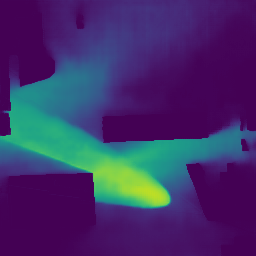}
    \caption*{0.0060}
    \end{subfigure}
    \begin{subfigure}[t]{0.033\textwidth}
        \includegraphics[width=\textwidth]{jaens5}
    \end{subfigure}
    \\
    \begin{subfigure}[t]{0.12\textwidth}
    \includegraphics[width=\textwidth]{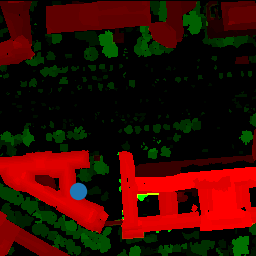}
    \caption{}
    \label{fig:comparison_models_different:gis}
    \end{subfigure}
    \begin{subfigure}[t]{0.12\textwidth}
    \includegraphics[width=\textwidth]{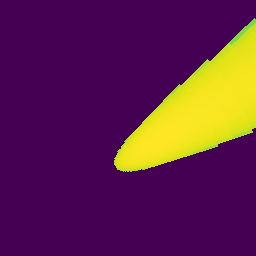}
    \caption{}
    \label{fig:comparison_models_different:gain}
    \end{subfigure}
    \begin{subfigure}[t]{0.033\textwidth}
        \includegraphics[width=\textwidth]{jaens25}
    \end{subfigure}
    \begin{subfigure}[t]{0.12\textwidth}
    \includegraphics[width=\textwidth]{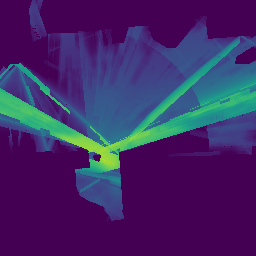}
    \caption{}
    \label{fig:comparison_models_different:target}
    \end{subfigure}
    \begin{subfigure}[t]{0.12\textwidth}
    \includegraphics[width=\textwidth]{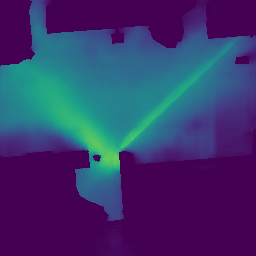}
    \caption{0.0280}
    \label{fig:comparison_models_different:radiounet}
    \end{subfigure}
    \begin{subfigure}[t]{0.12\textwidth}
    \includegraphics[width=\textwidth]{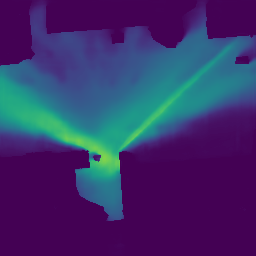}
    \caption{0.0162}
    \label{fig:comparison_models_different:pmnet}
    \end{subfigure}
    \begin{subfigure}[t]{0.12\textwidth}
    \includegraphics[width=\textwidth]{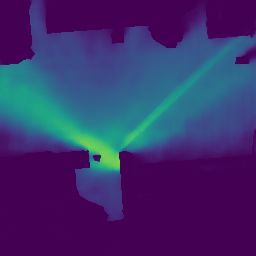}
    \caption{0.0142}
    \label{fig:comparison_models_different:dcn}
    \end{subfigure}    
    \begin{subfigure}[t]{0.033\textwidth}
        \includegraphics[width=\textwidth]{jaens5}
    \end{subfigure}
    \caption{Samples with clear visual differences between the models -- 
    \subref{fig:comparison_models_different:gis} overlayed height maps with buildings in red, vegetation in green and Tx position in blue 
    \subref{fig:comparison_models_different:gain} antenna gain projected onto the floor 
    \subref{fig:comparison_models_different:target} ground truth radio map
    predictions by the best versions (Table \ref{table:geometry_los}) of: 
        \subref{fig:comparison_models_different:radiounet} RadioUNet \cite{radiounet} 
        \subref{fig:comparison_models_different:pmnet} PMNet \cite{pmnet} 
        \subref{fig:comparison_models_different:dcn} UNetDCN, MSE below.
    }
    \label{fig:comparison_models_different}
\end{figure*}

To test the effectiveness of the previously discussed input features, we have conducted extensive experiments with the three models described in Section \ref{sec:architectures}.
In all experiments, unless stated otherwise, the networks receive the one-hot encoding of the Tx location, building and vegetation nDSMs and the antenna gain projected onto the floor and building top (\textit{basic}) and potentially additional inputs.
The results for different city geometry encodings and LoS information are presented in Table \ref{table:geometry_los}.
It can be seen that without additional LoS information, all of the discussed city geometry encodings provide a solid improvement over the the baseline inputs, with the only exception being the spherical coordinate system.
Adding in LoS information, the differences become smaller, although the geometry encodings still provide some additional benefit.
This makes sense as one key idea of GA and the relative coordinate systems is helping the network to determine the direct paths from the Tx to different positions on the map.
One can observe from Fig. \ref{fig:comparison_models_different} however, that they are  also helpful to recognize reflections in some cases.

Our proposed method of encoding the LoS information provides a slight improvement compared to the binary encoding, with again the exception of the spherical coordinates.
The RadioUNet benefits less from the additional inputs and performs worse than the other two models in general, which we attribute to the significantly lower complexity and the lack of any enhancements to aid with the propagation of information over longer distances.
\begin{figure*}[!htb]
    \centering
    \begin{subfigure}[t]{0.12\textwidth}
    \includegraphics[width=\textwidth]{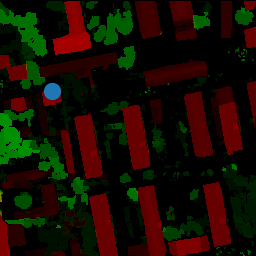}
    \end{subfigure}
    \begin{subfigure}[t]{0.12\textwidth}
    \includegraphics[width=\textwidth]{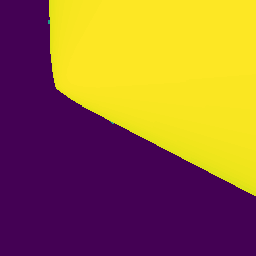}
    \end{subfigure}
    \begin{subfigure}[t]{0.033\textwidth}
        \includegraphics[width=\textwidth]{jaens25}
    \end{subfigure}
    \begin{subfigure}[t]{0.12\textwidth}
    \includegraphics[width=\textwidth]{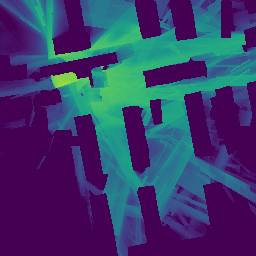}
    \end{subfigure}
    \begin{subfigure}[t]{0.12\textwidth}
    \includegraphics[width=\textwidth]{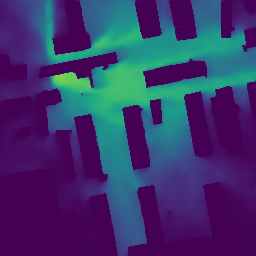}
    \caption*{0.00612}
    \end{subfigure}
    \begin{subfigure}[t]{0.12\textwidth}
    \includegraphics[width=\textwidth]{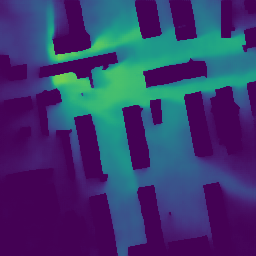}
    \caption*{0.00480}
    \end{subfigure}
    \begin{subfigure}[t]{0.12\textwidth}
    \includegraphics[width=\textwidth]{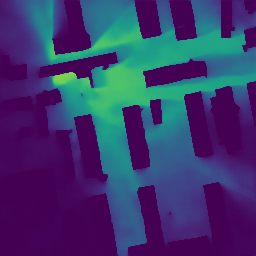}
    \caption*{0.00437}
    \end{subfigure}
    \begin{subfigure}[t]{0.033\textwidth}
        \includegraphics[width=\textwidth]{jaens5}
    \end{subfigure}
    \\
    \begin{subfigure}[t]{0.12\textwidth}
    \includegraphics[width=\textwidth]{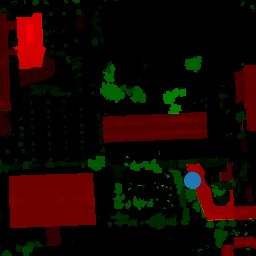}
    \end{subfigure}
    \begin{subfigure}[t]{0.12\textwidth}
    \includegraphics[width=\textwidth]{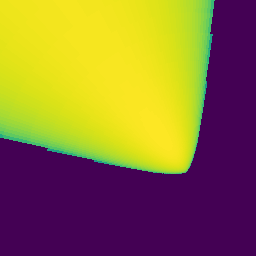}
    \end{subfigure}
    \begin{subfigure}[t]{0.033\textwidth}
        \includegraphics[width=\textwidth]{jaens25}
    \end{subfigure}
    \begin{subfigure}[t]{0.12\textwidth}
    \includegraphics[width=\textwidth]{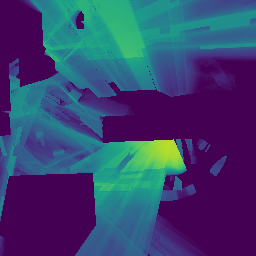}
    \end{subfigure}
    \begin{subfigure}[t]{0.12\textwidth}
    \includegraphics[width=\textwidth]{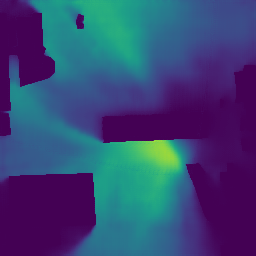}
    \caption*{0.01017}
    \end{subfigure}
    \begin{subfigure}[t]{0.12\textwidth}
    \includegraphics[width=\textwidth]{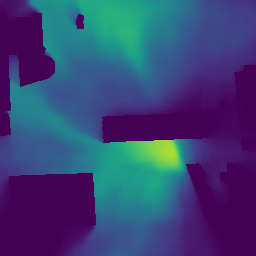}
    \caption*{0.00850}
    \end{subfigure}
    \begin{subfigure}[t]{0.12\textwidth}
    \includegraphics[width=\textwidth]{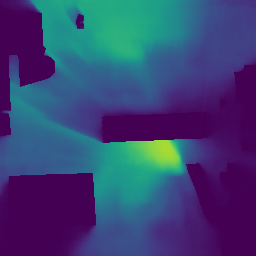}
    \caption*{0.00745}
    \end{subfigure}
    \begin{subfigure}[t]{0.033\textwidth}
        \includegraphics[width=\textwidth]{jaens5}
    \end{subfigure}
    \\
    \begin{subfigure}[t]{0.12\textwidth}
    \includegraphics[width=\textwidth]{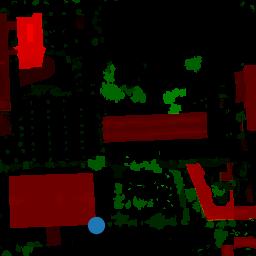}
    \end{subfigure}
    \begin{subfigure}[t]{0.12\textwidth}
    \includegraphics[width=\textwidth]{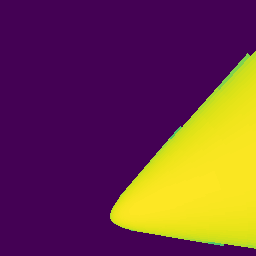}
    \end{subfigure}
    \begin{subfigure}[t]{0.033\textwidth}
        \includegraphics[width=\textwidth]{jaens25}
    \end{subfigure}
    \begin{subfigure}[t]{0.12\textwidth}
    \includegraphics[width=\textwidth]{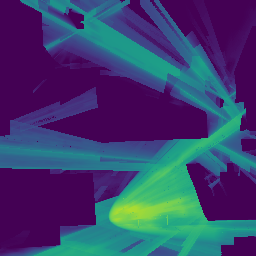}
    \end{subfigure}
    \begin{subfigure}[t]{0.12\textwidth}
    \includegraphics[width=\textwidth]{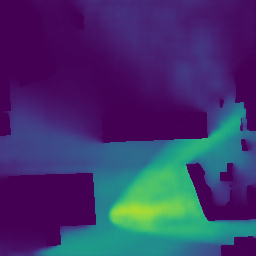}
    \caption*{0.02055}
    \end{subfigure}
    \begin{subfigure}[t]{0.12\textwidth}
    \includegraphics[width=\textwidth]{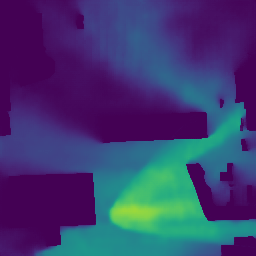}
    \caption*{0.01392}
    \end{subfigure}
    \begin{subfigure}[t]{0.12\textwidth}
    \includegraphics[width=\textwidth]{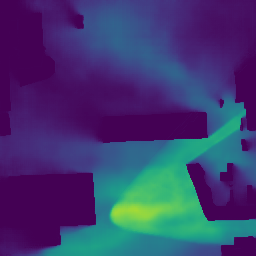}
    \caption*{0.01381}
    \end{subfigure}
    \begin{subfigure}[t]{0.033\textwidth}
        \includegraphics[width=\textwidth]{jaens5}
    \end{subfigure}
    \\
    \begin{subfigure}[t]{0.12\textwidth}
    \includegraphics[width=\textwidth]{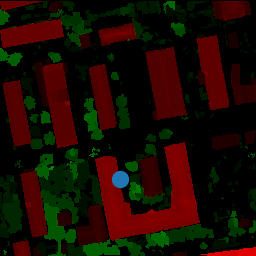}
    \caption{}
    \label{fig:comparison_inputs:gis}
    \end{subfigure}
    \begin{subfigure}[t]{0.12\textwidth}
    \includegraphics[width=\textwidth]{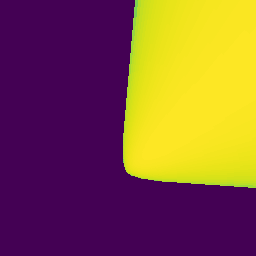}
    \caption{}
    \label{fig:comparison_inputs:gain}
    \end{subfigure}
    \begin{subfigure}[t]{0.033\textwidth}
        \includegraphics[width=\textwidth]{jaens25}
    \end{subfigure}
    \begin{subfigure}[t]{0.12\textwidth}
    \includegraphics[width=\textwidth]{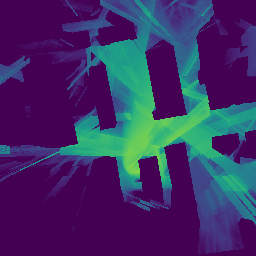}
    \caption{}
    \label{fig:comparison_inputs:target}
    \end{subfigure}
    \begin{subfigure}[t]{0.12\textwidth}
    \includegraphics[width=\textwidth]{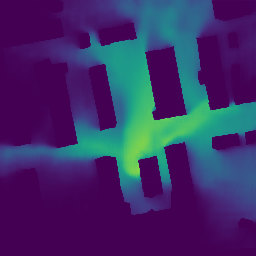}
    \caption{0.00765}
    \label{fig:comparison_inputs:basic}
    \end{subfigure}
    \begin{subfigure}[t]{0.12\textwidth}
    \includegraphics[width=\textwidth]{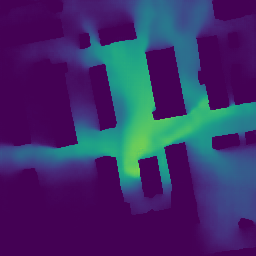}
    \caption{0.00710}
    \label{fig:comparison_inputs:coords}
    \end{subfigure}
    \begin{subfigure}[t]{0.12\textwidth}
    \includegraphics[width=\textwidth]{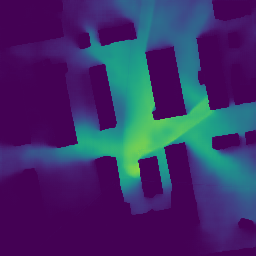}
    \caption{0.00527}
    \label{fig:comparison_inputs:coordslos}
    \end{subfigure}    
    \begin{subfigure}[t]{0.033\textwidth}
        \includegraphics[width=\textwidth]{jaens5}
    \end{subfigure}
    \caption{Samples with clear visual differences between the inputs -- 
    \subref{fig:comparison_inputs:gis} overlayed height maps with buildings in red, vegetation in green and Tx position in blue 
    \subref{fig:comparison_inputs:gain} antenna gain projected onto the floor 
    \subref{fig:comparison_inputs:target} ground truth radio map, predictions by UNetDCN with: 
        \subref{fig:comparison_inputs:basic} baseline inputs 
        \subref{fig:comparison_inputs:coords} cylindrical coordinates 
        \subref{fig:comparison_inputs:coordslos} cylindrical coordinates and LoS information, MSE below. }
    \label{fig:comparison_inputs}
\end{figure*}

\begin{figure*}[!htb]
    \centering
    \begin{subfigure}[t]{0.12\textwidth}
    \includegraphics[width=\textwidth]{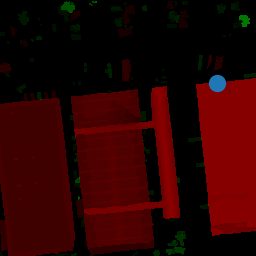}
    \end{subfigure}
    \begin{subfigure}[t]{0.12\textwidth}
    \includegraphics[width=\textwidth]{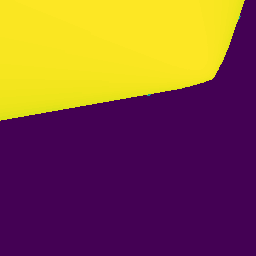}
    \end{subfigure}
    \begin{subfigure}[t]{0.033\textwidth}
        \includegraphics[width=\textwidth]{jaens25}
    \end{subfigure}
    \begin{subfigure}[t]{0.12\textwidth}
    \includegraphics[width=\textwidth]{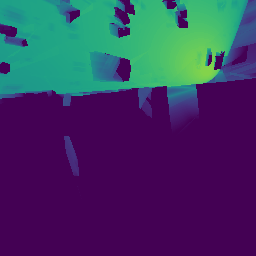}
    \end{subfigure}
    \begin{subfigure}[t]{0.12\textwidth}
    \includegraphics[width=\textwidth]{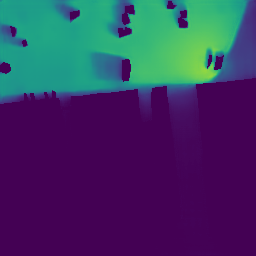}
    \caption*{0.00234}
    \end{subfigure}
    \begin{subfigure}[t]{0.12\textwidth}
    \includegraphics[width=\textwidth]{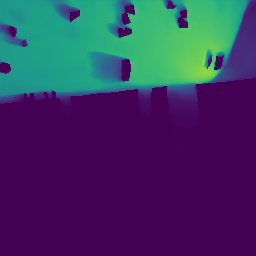}
    \caption*{0.00215}
    \end{subfigure}
    \begin{subfigure}[t]{0.12\textwidth}
    \includegraphics[width=\textwidth]{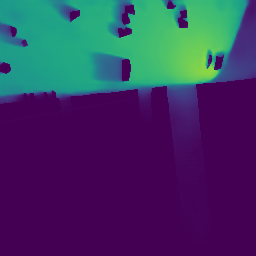}
    \caption*{0.00204}
    \end{subfigure}
    \begin{subfigure}[t]{0.033\textwidth}
        \includegraphics[width=\textwidth]{jaens5}
    \end{subfigure}
    \\
    \begin{subfigure}[t]{0.12\textwidth}
    \includegraphics[width=\textwidth]{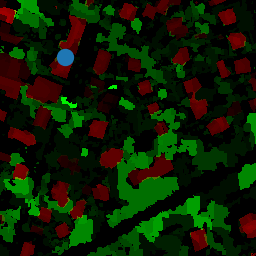}
    \end{subfigure}
    \begin{subfigure}[t]{0.12\textwidth}
    \includegraphics[width=\textwidth]{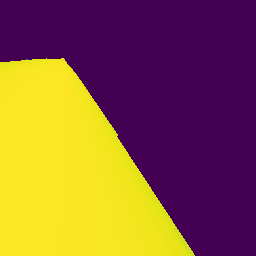}
    \end{subfigure}
    \begin{subfigure}[t]{0.033\textwidth}
        \includegraphics[width=\textwidth]{jaens25}
    \end{subfigure}
    \begin{subfigure}[t]{0.12\textwidth}
    \includegraphics[width=\textwidth]{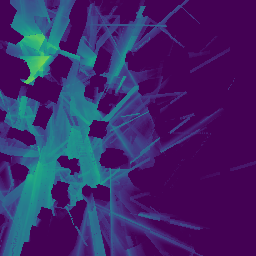}
    \end{subfigure}
    \begin{subfigure}[t]{0.12\textwidth}
    \includegraphics[width=\textwidth]{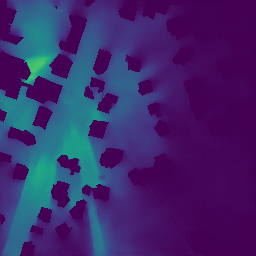}
    \caption*{0.00574}
    \end{subfigure}
    \begin{subfigure}[t]{0.12\textwidth}
    \includegraphics[width=\textwidth]{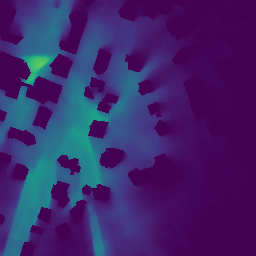}
    \caption*{0.00661}
    \end{subfigure}
    \begin{subfigure}[t]{0.12\textwidth}
    \includegraphics[width=\textwidth]{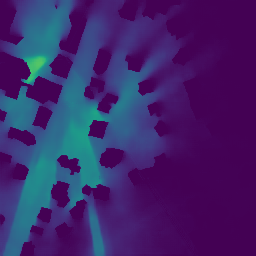}
    \caption*{0.00643}
    \end{subfigure}
    \begin{subfigure}[t]{0.033\textwidth}
        \includegraphics[width=\textwidth]{jaens5}
    \end{subfigure}
    \\
    \begin{subfigure}[t]{0.12\textwidth}
    \includegraphics[width=\textwidth]{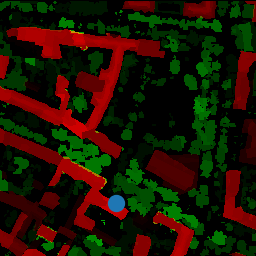}
    \end{subfigure}
    \begin{subfigure}[t]{0.12\textwidth}
    \includegraphics[width=\textwidth]{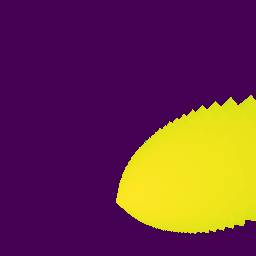}
    \end{subfigure}
    \begin{subfigure}[t]{0.033\textwidth}
        \includegraphics[width=\textwidth]{jaens25}
    \end{subfigure}
    \begin{subfigure}[t]{0.12\textwidth}
    \includegraphics[width=\textwidth]{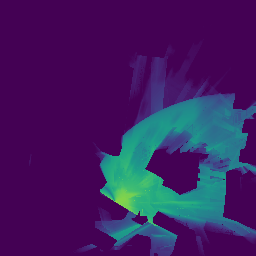}
    \end{subfigure}
    \begin{subfigure}[t]{0.12\textwidth}
    \includegraphics[width=\textwidth]{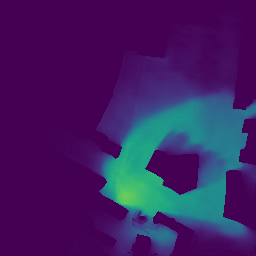}
    \caption*{0.00291}
    \end{subfigure}
    \begin{subfigure}[t]{0.12\textwidth}
    \includegraphics[width=\textwidth]{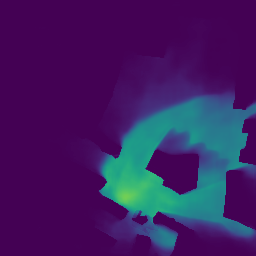}
    \caption*{0.00226}
    \end{subfigure}
    \begin{subfigure}[t]{0.12\textwidth}
    \includegraphics[width=\textwidth]{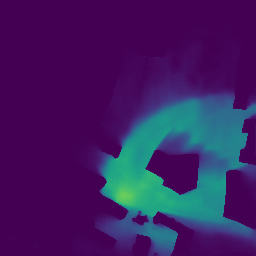}
    \caption*{0.00205}
    \end{subfigure}
    \begin{subfigure}[t]{0.033\textwidth}
        \includegraphics[width=\textwidth]{jaens5}
    \end{subfigure}
    \\
    \begin{subfigure}[t]{0.12\textwidth}
    \includegraphics[width=\textwidth]{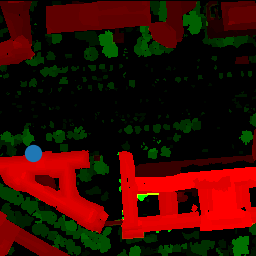}
    \caption{}
    \label{fig:comparison_models_similar:gis}
    \end{subfigure}
    \begin{subfigure}[t]{0.12\textwidth}
    \includegraphics[width=\textwidth]{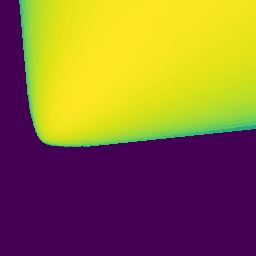}
    \caption{}
    \label{fig:comparison_models_similar:gain}
    \end{subfigure}
    \begin{subfigure}[t]{0.033\textwidth}
        \includegraphics[width=\textwidth]{jaens25}
    \end{subfigure}
    \begin{subfigure}[t]{0.12\textwidth}
    \includegraphics[width=\textwidth]{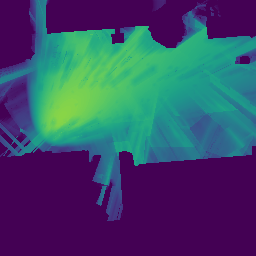}
    \caption{}
    \label{fig:comparison_models_similar:target}
    \end{subfigure}
    \begin{subfigure}[t]{0.12\textwidth}
    \includegraphics[width=\textwidth]{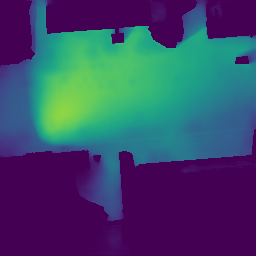}
    \caption{0.00358}
    \label{fig:comparison_models_similar:radiounet}
    \end{subfigure}
    \begin{subfigure}[t]{0.12\textwidth}
    \includegraphics[width=\textwidth]{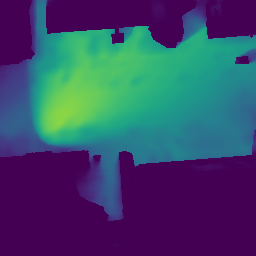}
    \caption{0.00286}
    \label{fig:comparison_models_similar:pmnet}
    \end{subfigure}
    \begin{subfigure}[t]{0.12\textwidth}
    \includegraphics[width=\textwidth]{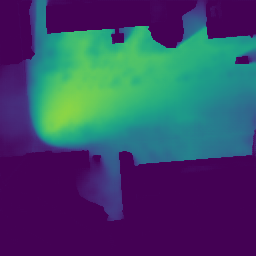}
    \caption{0.00355}
    \label{fig:comparison_models_similar:dcn}
    \end{subfigure}    
    \begin{subfigure}[t]{0.033\textwidth}
        \includegraphics[width=\textwidth]{jaens5}
    \end{subfigure}
    \caption{Samples with very similar estimates by all models --
    \subref{fig:comparison_models_similar:gis} overlayed height maps with buildings in red, vegetation in green and Tx position in blue,
    \subref{fig:comparison_models_similar:gain} antenna gain projected onto the floor,
    \subref{fig:comparison_models_similar:target} ground truth radio map, predictions by the best versions (Table \ref{table:geometry_los}) of: 
        \subref{fig:comparison_models_similar:radiounet} RadioUNet \cite{radiounet},
        \subref{fig:comparison_models_similar:pmnet} PMNet \cite{pmnet},
        \subref{fig:comparison_models_similar:dcn} UNetDCN, MSE below.}
    \label{fig:comparison_models_similar}
\end{figure*}

The numeric errors appear very close. 
Visual inspection of the predictions shows that we have very different types of samples in our dataset, on some of these we can see clear {differences} between models and/or inputs, whereas other predictions are {very similar} visually and in terms of loss.
With the city geometry encodings and LoS information, all models predict path loss inside the LoS areas fairly well, but we can see in Fig. \ref{fig:comparison_models_different} that the RadioUNet may struggle more than the other models to propagate the signal over a long distance after a reflection.
It also tends to blur larger areas in some cases.
On the other hand, when the relevant part of the radio map coincides with the area the Tx is pointing at, the models produce rather similar output, see Fig. \ref{fig:comparison_models_similar}.

In Table \ref{table:ant_encoding} we list the performances with varying antenna encodings explained in Section \ref{sec:antennas}.
In this case, the differences are very small and not consisted among different models. 
This suggests that there is no significant benefit from any of the tested alternatives to the baseline.

\begin{table*}[!htb]
    \centering
    \begin{tabular}{|c||c|c|c|c|c|c|}
        \hline
        Model  &   \multicolumn{2}{c|}{RadioUNet \cite{radiounet}}  &   \multicolumn{2}{c|}{PMNet \cite{pmnet}}                                      &   \multicolumn{2}{c|}{UNetDCN}  \\
        \hline
        \hline
        Encoding                &   RMSE            &   NMSE              &   RMSE                  &   NMSE            &   RMSE                &   NMSE    \\
        \hline
        gain floor              &  $0.0718$         &  $0.00219$          &   $0.0655$            &  ${0.00182}$        &  $\textbf{0.0653}$    &  $\textbf{0.00181}$  \\
        gain floor + top        &  $0.0713$         &  $0.00217$          &   $0.0658$            &  ${0.00184}$        &  $\textbf{0.0645}$    &  $\textbf{0.00178}$      \\
        \hline
        FSPL floor              &  $0.0724$         &  $0.00223$          &  ${0.0655}$    &  ${{0.00182}}$ &   \textbf{${0.0646}$ }& \textbf{${0.00177}$  }\\
        FSPL floor + top        &  $\textit{0.0701}$&  $\textit{0.00208}$ &  $0.0650$             &  $0.00179$          &   $\textbf{0.0645}$    &  $\textbf{0.00177}$  \\  
        \hline
        slices ($4$m)           &  $0.0727$         &  $0.00226$          &  $\textit{0.0644}$    &   $\textit{0.00176}$&   $\textbf{{0.0642}}$&  $\textbf{{0.00175}}$  \\
        slices ($1$m)           &  $0.0724$         &  $0.00224$          &  $0.0648$             &  $0.00178$          &   $\textbf{\textit{0.0639}}$    &   $\textbf{\textit{0.00173}}$  \\
        FSPL slices ($4$m)      &  $0.0722$         &  ${{0.00223}}$      &$\textit{\textbf{0.0644}}$&  $\textbf{\textit{0.00176}}$&   ${{0.0648}}$&   ${{0.00179}}$ \\
        \hline
    \end{tabular}
    \caption{Antenna encoding -- lowest errors in each column (input) in \textit{italics} and per row (model) in \textbf{bold}.}
    \label{table:ant_encoding}
\end{table*}

As described in the introduction, precise information about the locations, heights and shapes of buildings and especially vegetation are scarce.
Aerial imagery on the other hand is an, in many cases, easier accessible data source which at least partially contains this information implicitly.
We perform further experiments to see to what extend the models are capable of predicting the radio map from just images or potentially with additional sight information (see Table \ref{table:img}).
Intuitively, this requires the models to implicitly perform a semantic segmentation, i.e. pixel-wise classification, of the input image in order to find objects relevant for the signal propagation, and to estimate the heights of the found objects.
One additional input feature ((\textit{coords} in Table \ref{table:img})) we consider is a combination of GA with the azimuth angle and 2D distance from the cylindrical coordinate system considered before, as these are solely tied to the Tx position and the two-dimensional locations on the map, but not to the nDSMs.
The images contain, besides the usual RGB channels, an infrared channel, which we include by default. 
Performance without it is also tested (\textit{w/o IR}).
Lastly, we additionally provide the networks with an unclassified {nDSM} depicting elevation of buildings and vegetation together, as this kind of information is easier to acquire than height maps for each class individually.

\begin{figure*}[!htb]
    \centering
    \begin{subfigure}[t]{0.12\textwidth}
    \includegraphics[width=\textwidth]{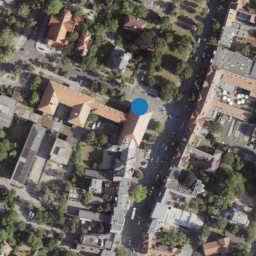}
    \end{subfigure}
    \begin{subfigure}[t]{0.12\textwidth}
    \includegraphics[width=\textwidth]{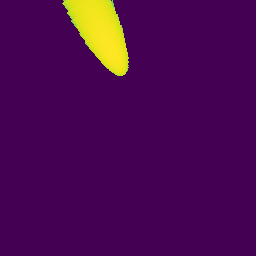}
    \end{subfigure}
    \begin{subfigure}[t]{0.033\textwidth}
        \includegraphics[width=\textwidth]{jaens25}
    \end{subfigure}
    \begin{subfigure}[t]{0.12\textwidth}
    \includegraphics[width=\textwidth]{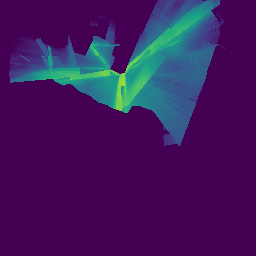}
    \end{subfigure}
    \begin{subfigure}[t]{0.12\textwidth}
    \includegraphics[width=\textwidth]{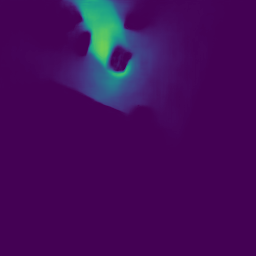}
    \caption*{0.03284}
    \end{subfigure}
    \begin{subfigure}[t]{0.12\textwidth}
    \includegraphics[width=\textwidth]{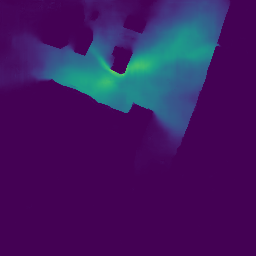}
    \caption*{0.00492}
    \end{subfigure}
    \begin{subfigure}[t]{0.033\textwidth}
        \includegraphics[width=\textwidth]{jaens5}
    \end{subfigure}
    \\
    \begin{subfigure}[t]{0.12\textwidth}
    \includegraphics[width=\textwidth]{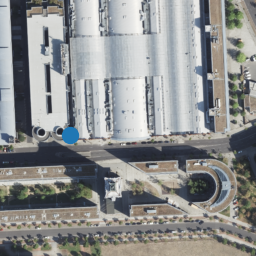}
    \end{subfigure}
    \begin{subfigure}[t]{0.12\textwidth}
    \includegraphics[width=\textwidth]{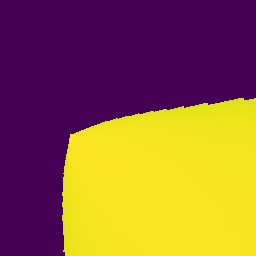}
    \end{subfigure}
    \begin{subfigure}[t]{0.033\textwidth}
        \includegraphics[width=\textwidth]{jaens25}
    \end{subfigure}
    \begin{subfigure}[t]{0.12\textwidth}
    \includegraphics[width=\textwidth]{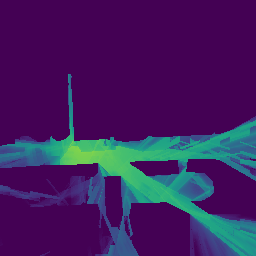}
    \end{subfigure}
    \begin{subfigure}[t]{0.12\textwidth}
    \includegraphics[width=\textwidth]{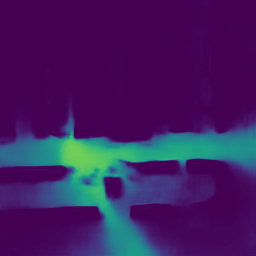}
    \caption*{0.02741}
    \end{subfigure}
    \begin{subfigure}[t]{0.12\textwidth}
    \includegraphics[width=\textwidth]{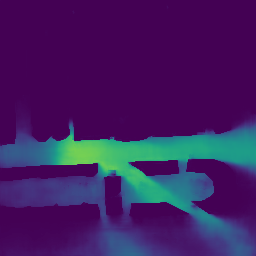}
    \caption*{0.00620}
    \end{subfigure}
    \begin{subfigure}[t]{0.033\textwidth}
        \includegraphics[width=\textwidth]{jaens5}
    \end{subfigure}
    \\
    \begin{subfigure}[t]{0.12\textwidth}
    \includegraphics[width=\textwidth]{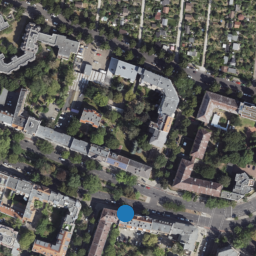}
    \end{subfigure}
    \begin{subfigure}[t]{0.12\textwidth}
    \includegraphics[width=\textwidth]{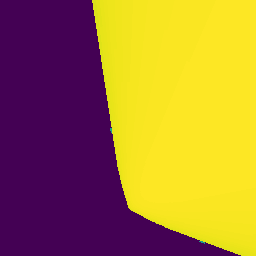}
    \end{subfigure}
    \begin{subfigure}[t]{0.033\textwidth}
        \includegraphics[width=\textwidth]{jaens25}
    \end{subfigure}
    \begin{subfigure}[t]{0.12\textwidth}
    \includegraphics[width=\textwidth]{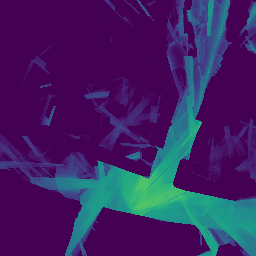}
    \end{subfigure}
    \begin{subfigure}[t]{0.12\textwidth}
    \includegraphics[width=\textwidth]{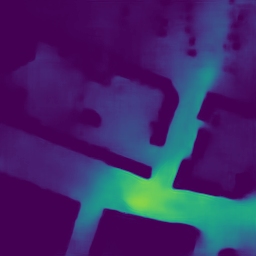}
    \caption*{0.00827}
    \end{subfigure}
    \begin{subfigure}[t]{0.12\textwidth}
    \includegraphics[width=\textwidth]{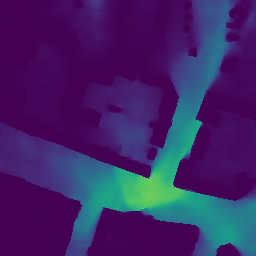}
    \caption*{0.00602}
    \end{subfigure}
    \begin{subfigure}[t]{0.033\textwidth}
        \includegraphics[width=\textwidth]{jaens5}
    \end{subfigure}
    \\
    \begin{subfigure}[t]{0.12\textwidth}
    \includegraphics[width=\textwidth]{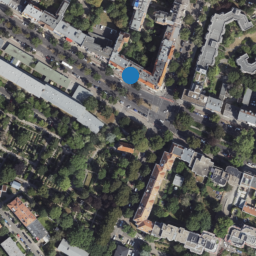}
    \caption{}
    \label{fig:comparison_images:img}
    \end{subfigure}
    \begin{subfigure}[t]{0.12\textwidth}
    \includegraphics[width=\textwidth]{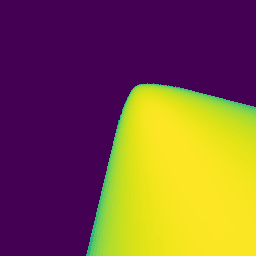}
    \caption{}
    \label{fig:comparison_images:gain}
    \end{subfigure}
    \begin{subfigure}[t]{0.033\textwidth}
        \includegraphics[width=\textwidth]{jaens25}
    \end{subfigure}
    \begin{subfigure}[t]{0.12\textwidth}
    \includegraphics[width=\textwidth]{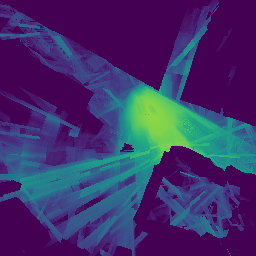}
    \caption{}
    \label{fig:comparison_images:target}
    \end{subfigure}
    \begin{subfigure}[t]{0.12\textwidth}
    \includegraphics[width=\textwidth]{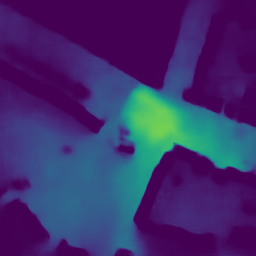}
    \caption{0.00963}
    \label{fig:comparison_images:predimg}
    \end{subfigure}
    \begin{subfigure}[t]{0.12\textwidth}
    \includegraphics[width=\textwidth]{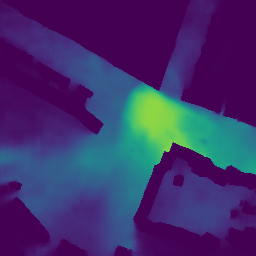}
    \caption{0.00715}
    \label{fig:comparison_images:predimgndsm}
    \end{subfigure}    
    \begin{subfigure}[t]{0.033\textwidth}
        \includegraphics[width=\textwidth]{jaens5}
    \end{subfigure}
    \caption{An example of prediction from image and potentially nDSM. 
    \subref{fig:comparison_images:img} aerial image and Tx position in blue 
    \subref{fig:comparison_images:gain} antenna gain projected onto the floor 
    \subref{fig:comparison_images:target} ground truth radio map, 
    predictions by PMNet \cite{pmnet} with inputs: 
        \subref{fig:comparison_images:predimg} only image 
        \subref{fig:comparison_images:predimgndsm} image and unclassified nDSM, MSE below.}
    \label{fig:comparison_images}
\end{figure*}

In Table \ref{table:img} we show the results the models achieve in these settings.
The performance is significantly worse if height information is not available, which is to be expected.
Some comparisons are shown in Fig. \ref{fig:comparison_images}.
The position and approximate shape of buildings and trees seem to be recognized  relatively well by the networks, but inferring the height of objects from a two-dimensional image is an inherently ill-posed problem \cite{im2height} and does not always work well.
In the first row (Fig. \ref{fig:comparison_images}) for example, we can see that the network without access to height information does not recognize that the house in the main lobe completely blocks the signal.
Below in the second row, it seems to misinterpret the exact locations of the buildings and consequently predicts a non-existent propagation path.
The prediction of the other two samples shown is satisfactory though.
As mentioned in Section \ref{sec:related_works}, the idea of estimating the path loss from aerial or satellite images is not completely new \cite{plgan}.
This is, however, the first attempt to predict the complete radio map for the challenging setting of urban cellular environments with Tx placed on buildings, as far as we are aware.

\begin{table*}
    \centering
    \begin{tabular}{|c||c|c|c|c|c|c|}
        \hline
        Model                  &   \multicolumn{2}{c|}{RadioUNet \cite{radiounet}}  &   \multicolumn{2}{c|}{PMNet \cite{pmnet}}         &   \multicolumn{2}{c|}{UNetDCN}   \\
        \hline
        \hline
        Input                   &   RMSE                &   NMSE                    &   RMSE                    &   NMSE                &   RMSE                    &   NMSE \\
        \hline
        Image                   &   $\textit{0.0940}$   &   $\textit{0.00378}$      &   $\textit{0.0914}$       &  $\textit{0.00359}$   &   $\textbf{0.0885}$       &   $\textbf{0.00335}$ \\
        Image w/o IR            &   $0.0968$            &   $0.00402$               &   $0.0931$                &  $0.00371$            &   $\textbf{\textit{0.0873}}$&   $\textbf{\textit{0.00327}}$   \\ 
        Image + coords          &   $0.0980$            &   $0.00412$               &   $0.0917$                &  $0.00362$            &   $\textbf{0.0888}$       &   $\textbf{0.00338}$   \\
        \hline
        Image + nDSM            &   $0.0758$            &   $0.00245$               &   $\textbf{0.0702}$       &  $\textbf{0.00210}$   &   $0.0721$                &   $0.00221$ \\
        Image + nDSM + coords   &   $\underline{0.0733}$&   $\underline{0.00228}$   &   $\textbf{\underline{0.0680}}$&  $\textbf{\underline{0.00196}}$&   $\underline{0.0685}$&   $\underline{0.00199}$ \\
        \hline
    \end{tabular}
    \caption{Prediction from images -- lowest errors in each column (input) without nDSM in \textit{italics}, with nDSM \underline{underlined} and per row (model) in \textbf{bold}.}
    \label{table:img}
\end{table*}

\section{CONCLUSION}
We hope that by making our dataset and code available for public use, we facilitate the work of other researchers and contribute to more reproducible and comparable investigations in the field.
Featuring directive Tx antennas, the dataset opens potential for the investigation of down-stream tasks in 5G/6G networks such as beam codebook design and beam management \cite{beams}.

The possibility to train models to predict the radio map from only aerial imagery is an exciting aspect which deserves further attention. 
Since this data source is very different from already classified height maps or binary class images, new model architectures or different hyperparameters may possibly be more suited for this case and reach higher accuracies than the models considered in this work.
As a further potential extension of this, predicting the radio map from satellite data available for the whole planet (similarly to \cite{plgan}, but in cellular networks) would allow large-scale applications to network planning and related tasks.
The often lower spatial resolution may make the recognition of objects and shapes more difficult.
On the other hand, incorporation of other spectral bands or radar data could be beneficial for the implicit classification of objects and height estimation \cite{sentinel_height}.

While this is out of the scope of this work, the dataset could also serve to investigate the problem of joint semantic segmentation and height estimation from aerial images \cite{im2height}, \cite{joint_seg_height}, which in turn could be an intermediate step in the prediction of the radio map.

\section*{ACKNOWLEDGMENT}
We thank \c{C}a\v{g}kan Yapar for advice on the dataset generation and the related literature, Ron Levie for discussions on how to generate and use line-of-sight information, Tom Burgert for advice on curriculum learning and the recommendation to use PyTorch Lightning and Saeid Dehkordi for proofreading.

\clearpage
\printbibliography

\end{document}